\documentclass[12 pt]{article}
\usepackage{graphicx}
\begin{document}
\baselineskip=24pt
\title{\bf{DGLAP Evolutions and cross-sections of Neutrino-Nucleon interaction at Ultra High energy }}
\author{D K Choudhury \thanks{E-mail: dkc.physics@gmail.com} \\Department of Physics\\Gauhati University\\Guwahati-781014, India \and  Pijush Kanti Dhar \thanks{E-mail: dr\_pijush@yahoo.co.in}\\Department of Physics\\Arya Vidyapeeth College\\Guwahati-781016, India}
\maketitle

\begin{abstract}
We determine the UHE neutrino-nucleon cross-sections analytically both for charged and neutral current in the leading order by using the solutions of non-singlet DGLAP equation as well as singlet DGLAP equation respectively. Next we determine the cross-sections for both charged and neutral current at one loop level by carring out numerical integration of the double differential cross-section.  For analytical solution, we take the standard MRST 2004 f4 LO parton distributions  at $Q_0^2=1$ \,GeV$^2$ as input and for numerical solution (exact), we consider the GRV 94 parton distributions. 

Our analytical as well as numerically determined results are compared with the NLO results of various authors. We find our numerically determined result in sufficient agreement with the results obtained by those authors. On the other hand, our analytical result matches better with other results at lower end of the energy spectrum of the UHE neutrino rather than the higher end. 



\end{abstract}

\section{Introduction}

Charged current induced interactions in the Ultra-High Energy (UHE) neutrino (antineutrino)-nucleon scattering processes are represented as

$\nu_\mu N\rightarrow\mu^{-}X$,\,\,${\overline\nu_\mu}N\rightarrow \mu^{+}X$, where $N=(n+p)/2$ is an isoscaler nucleon.

On the other hand, neutral current interactions are represented as:

$\nu_\mu N\rightarrow\nu_\mu X$,\,\,${\overline\nu_\mu}N\rightarrow{\overline\nu_\mu}X$

In case of charged current interactions, the mediating particles are W-bosons $(W^{\pm})$ and in case of neutral current interactions, the mediating particles are Z-bosons ($Z^0$).

Study of UHE energy neutrino-nucleon interaction on the basis of standard model of particle physics has been pursued by various authors \cite{RAHULBASU,RCMI1,GKR,RAMCFR,FIOREJKPPPR12,JJM,EMHJJM} from time to time since long back. Recent measurements indicate that there are GZK-violating air showers with reliably determined energies above $10^{11}$ GeV \cite{MTAKEDA, DJBIRD} triggered off by UHE neutrinos. Event rates of these showers can shoot up the UHE cross-sections several thousands times as predicted by standard model. This unusual bumping up of cross-sections for neutrinos having energies above $10^8$\, GeV gives rise to a puzzle  quite unexplainable by the standard model.

 UHE neutrinos having energies above $10^8$\, GeV collide violently with nucleons at center-of-mass energies \cite{INASARCEVIC} which is greater than

$\sqrt{S_{\nu N}}\equiv \sqrt{2ME_{\nu}}\simeq 14\,\displaystyle{\left(\frac{E_{\nu}}{10^{8}}\,\,\ GeV \right)}^{\frac{1}{2}}$\,\, TeV

At such  gigantic energies, the magnitudes of the ultrasmall Bjorken $x$ values are less than 

${x\simeq \displaystyle{2\times 10^{-4}\,\left(\frac{Q^2}{M_{W}^{2}}\right)\left(\frac{0.2}{y}\right)\left(\frac{10^8\,\,GeV}{E_{\nu}}\right)}}$

\begin{figure}[t]
\vspace{0.4in}
\begin{center}
\includegraphics[width=4in]{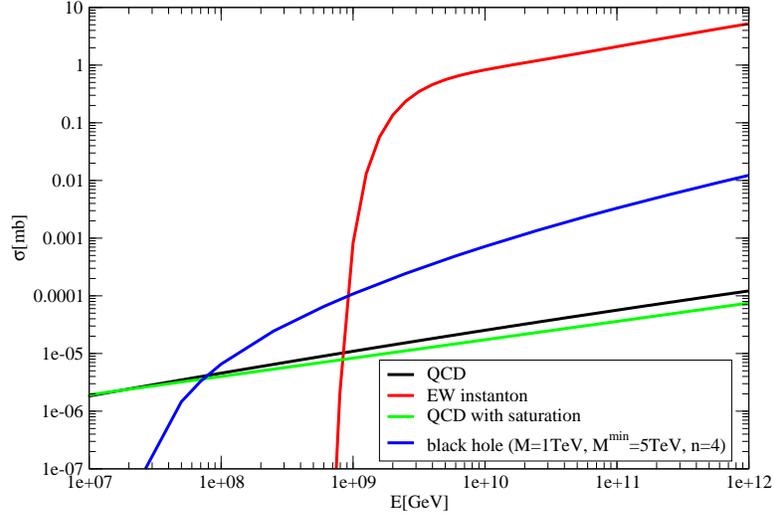}
\end{center}
\vspace{-0.1in}
\caption[Neutrino-nucleon interaction cross-section as per standard model and beyond.]{Neutrino-nucleon interaction cross-section as per standard model and beyond. Figure taken from Ref. \protect\cite{INASARCEVIC1145}} 
\label{fig:chapt1russia}
\end{figure}

At such high energies several approaches \cite{INASARCEVIC,INASARCEVIC1145} based on physics beyond the standard model have been proposed to explain the event rates measured by neutrino telescopes like 'Pierre Auger Observatory' \cite{JABRAHAMET,LAANCH,SEVOV,ABRAJHAM,ABRAJHAM12}, ICECUBE \cite{AKARLE,ICECUB,Michelangelo}, 'OWL' \cite{FWSTECKERKRBLMSS}, 'EUSO' \cite{eusorikeniasf} and 'JEMEUSO' \cite{GUSMEDTAN}. One such approach is based on  electroweak instanton-induced processes where there is interaction between cosmogenic neutrinos and the nucleons in the atmosphere. Such a process, which was proposed by Ringwald \cite{RINGWALDA} and Bezrukov et al \cite{FBEV,FBEV112} violates the baryon+lepton number, B + L. Other approaches include production of miniature black holes in ultra-relativistic collisions  with energies above the fundamental Planck scale $M_D$ $(M_D \sim TeV)$ \cite{FBEV112,ARHT,JLFADSH,LANGOLH,SIYERDMHRENOIS}, neutrino interactions through TeV string resonance processes with string scale $M_{S}$ \cite{HOOPERDTH}, production of quasi-stable charged sleptons (staus) predicted by weak scale supersymmetric models with the supersymmetry breaking scale greater than $5 \times 10^{6}$ GeV \cite{IAGBZC,NOICISSU,YHMHRISJUSC}. Stau, which has a decay length on the scale of 10 km, is the next-to-lightest particle (NLP) which is produced by neutrino interactions on earth or atmosphere. Detection of pairs of charged tracks of secondary particles produced by  staus in neutrino telescopes like 'ANITA' \cite{PMIO,PWGORHAM} would be an excellent way to probe the SUSY breaking scale. A relative comparison of neutrino-nucleon interaction cross-section as per standard model and beyond is observed in Figure \ref{fig:chapt1russia}.

In the unpolarized case, several authors had solved DGLAP evolution equations \cite {GRIBOVLIPA,LIPATOV,DOKSH,AP,GALTA} efficiently even up to NNLO accuracy with numerical technique. There are basically two approaches: one by using the Mellin tranform method (N space based) and the other by using the $x$ space method. The numerical efficiency of the first method is more than the second, but the second method is more flexible than the first because here one requires inputs only in $x$ space. Now-a-days, one can get several numerical packages  publicly  to solve the DGLAP evolution equations efficiently and accurately, e.g., PEGASUS \cite{VOAGT} , HOPPET \cite{GAVSAJURO,GAVSAJURO1}, QCDNUM \cite{MBOTJE}, CANDIA \cite{ACCCMG,ACCCMG11}. For example, one can use the Fortran package QCD-PEGASUS to perform the evolution using the symbolic moment-space solutions on a one-fits-all Mellin inversion contour to get a  flexible and very accurate solutions of the evolution equations. PEGASUS implies ‘Parton Evolution Generated Applying Symbolic U-matrix Solutions'. On the other hand, there are several high flexible $x$-space methods. For example, the HOPPET NNLO parton evolution package is a very good Fortran  package for carrying out DGLAP evolution and other common manipulations of PDFs. Similarly, the evolution equations for parton densities and fragmentation functions in perturbative QCD can have numerical solution with the help of the QCDNUM program, which is powerful to evove un-polarised parton densities upto NNLO in powers of the strong coupling constant $\alpha_{s}$. Precision Studies of the NNLO DGLAP Evolution at the LHC can be carried out with programme CANDIA (progrmming language C and Fortran)\cite{ACCCMG}. Similarly, NNLO Logarithmic Expansions and Precise Determinations of the Neutral Currents near the Z Resonance at the LHC (the Drell-Yan case) can be obtained using CANDIA \cite{ACCCMG11}. In spite of having these highly efficient numerical packages of solving DGLAP equations for the QCD evolution of parton distributions (PDFs), analytical approahes for solving DGLAP equations bear special interest. It is interesting to see whether the analytical solutions deviate at all from the exact numerical solutions and if so, to what extent.

 In recent years, an approximate method \cite{DKCJS,DKCJS1,JKSDKCGKM,ATRI,DKCATRI} to solve DGLAP equations \cite {GRIBOVLIPA,LIPATOV,DOKSH,AP,GALTA}  has been pursued with considerable phenomenological success. In that approach, we have used Taylor series expansion  to express these equations as partial differential equations in $x$ $\displaystyle{\left(x=\frac{Q^2}{2p.q}\right)}$ and $t$ $\displaystyle{\left(t=\log\frac{Q^2}{\Lambda^2}\right)}$, valid at low $x$.

The aim of the present paper is to apply these approximate solutions at low $x$ to the Standard Model Interactions of ultra-high energy neutrino with nucleon which has an important bearing in neutrino astronomy. In doing so, we assume that the approximate solutions of DGLAP equations are valid in the region of ultrasmall values of Bjorken $x$.  With this assumption, we will find leading order (LO) expressions for UHE cross-sections for charged as well as neutral neutrino-nucleon interaction ($\sigma_{CC},\sigma_{NC})$ within this formalism. All throughout our earlier papers \cite {DKCPKD,Campus} we have resorted to MRST distributions and obtained results on the basis of these distributions in non-singlet as well as singlet case. We are going to apply these results in the present paper, so here too we consider MRST distributions \cite{MRST2222,MRSTDURHAM} to determine our analytic results.

We will also find UHE neutrino-nucleon cross-sections by using the numerical solutions of DGLAP equations. For this purpose, we take the help of GRV94  parton distributions \cite{GRVARXIV55A}.

 There are several authors \cite{RAHULBASU,RCMI1,GKR,FIOREJKPPPR12,RCMI111,FIOREJKPPPR1} who has predicted the UHE neutrino-nucleon cross-sections by different techniques. We will then compare our predictions with the result of several such authors and study the range of validity. In Section \ref{secchap6:Formalism}, we outline the formalism while Section \ref{secchap6:Results and discussions} contains results and discussions and Section \ref{secchap6:Comments and conclusions} contains comments and conclusions.

\section{Formalism}
\label{secchap6:Formalism}
\subsection{Approximate solutions of DGLAP equations:}
\label{subsecchap6:Approximate solutions of DGLAP equations:}

The approximate solutions of DGLAP equations \cite {GRIBOVLIPA,LIPATOV,DOKSH,AP,GALTA} for non-singlet structure function $F^{NS}(x,t)$ and singlet structure function $F_2^{S}(x,t)$ are respectively given by \cite{DKCPKD,Campus}, 
 
\begin{equation}
\label{eqn:ch6eq1}
{F^{NS}(x,t)=F^{NS}(x,t_0)\left(\frac{t}{t_0}\right)^{n(x,t)} \;\;\;\;(n(1,t)>0,t \ge t_{0})}
\end{equation}
and
\begin{equation}
\label{eqn:ch6eq2}
{F_2^{S}(x,t)=F^{S}(x,t_0)\frac{t^{k(x,t)}}{t_0^{k(x,t_0)}}\frac{\left[X^{S}(x)\right]^{k(x,t)}}{\left[X^{S}(x)\right]^{k(x,t_0)}}} \;\;\;\; (k(1,t)>0,t \ge t_{0})     
\end{equation}
where $t=\log\left(\frac{Q^2}{\Lambda ^2}\right)$ and $t_0=\log\left(\frac{Q_0^2}{\Lambda ^2}\right)$, $Q^2(\equiv -q^2)$ is the four momentum transfer and $\Lambda ^2$ is the QCD cut-off parameter. 
$X^{S}(x)$ is given by \cite{DKCPKD} 
\begin{equation}
\label{eqn:ch5g}
{X^{S}(x)=\exp\left[-\int\frac{dx}{P_1(x)}\right]}
\end{equation}
where 
\begin{equation}
\label{eqn:Jagadish}
{P_1(x,t)=-A_f x\left[2\log\left(\frac{1}{x}\right)+(1-x^2)\right]}
\end{equation}
It is to be noted that in DIS neutrino scattering, $F^{NS}(x,t)$  is identified as $x F_3(x,t)$.

\subsection{UHE neutrino-nucleon cross-section}
\label{subsecchap6:UHE neutrino-nucleon cross-section:}
\subsubsection{Charged Current-Analytical Solution:}
\label{subsubsecchap6:Charged Current:}

The differential cross-section for the charged current (CC) UHE interaction is given by \cite{GKR},
\begin{eqnarray}
\label{eqn:ch6eq3}
\frac{d^2 \sigma_{CC}^{\nu(\overline{\nu})N}}{dxdy}&=&\frac{G_F^2 s}{2 \pi}\left(1+\frac{Q^2}{M_W^{2}}\right)^{-2} \left[(1-y)F_2^{{\nu(\overline{\nu})}^{CC}}(x,Q^2)+y^2 x F_1^{{\nu(\overline{\nu})}^{CC}}(x,Q^2)\right.\nonumber\\
& & \left. \pm y\left(1-\frac{y}{2}\right) x F_3^{{\nu(\overline{\nu})}^{CC}}(x,Q^2)\right]
\end{eqnarray}

where Fermi coupling constant $G_F=1.1663\times 10^{-5} GeV^{-2}$, $s=2ME_{\nu({\overline \nu})}$ is the centre-of-mass energy squared, $x$ is the usual Bjorken variable which represents fraction of the nucleon's momentum carried by the quark and $\displaystyle{y=\frac{\nu}{E_\nu}}$, where $\nu= E_{\nu}-E_{\mu}=E_{had}$. $E_{\nu}$ represents the energy carried by incoming neutrino, $E_{\mu}$ represents the energy carried by outgoing muon  and $E_{had}$ represents the energy carried by the final state. 

The total cross-section for UHE CC interaction in an isoscaler target $N [\equiv \frac{(p+n)}{2}]$ is given by,
\begin{equation}
{\sigma_{CC}^{\nu({\overline\nu})N}=\int_0^1 dx \int_0^1dy \frac{d^2 \sigma_{CC}^{\nu({\overline\nu})N}}{dxdy}}
 \end{equation}

Replacing $y$ by $Q^{2}$ in the L.H.S. of eq. \ref{eqn:ch6eq3}, we get the expression for the double differential cross-section for charged  current as

\begin{eqnarray}
\label{eqn:ch6eq4}
\frac{d^2 \sigma_{CC}^{\nu({\overline\nu})N}}{dxdQ^{2}}&=&\frac{G_F^2 }{4\pi}\left(1+\frac{Q^2}{M_W^2}\right)^{-2}\frac{1}{x} \left[Y_{+}F_{2}^{{\nu({\overline\nu})}^{CC}}(x,Q^2)\right.\nonumber \\
& &\left. -y^{2}F_{L}^{{\nu({\overline\nu})}^{CC}}(x,Q^{2})\pm Y_{-} x F_{3}^{{\nu({\overline\nu})}^{CC}}(x,Q^2)\right]
\end{eqnarray}

where
\begin{equation}
\label{eqn:ch6eq5}
{Y_{\pm}=1 \pm (1-y)^{2}}
\end{equation}
and
\begin{equation}
\label{eqn:ch6eq6}
{F_{L}^{CC}=F_{2}^{CC}-2xF_{1}^{CC}}
\end{equation}

The total cross-section for UHE CC interaction can be re-written as \cite{EMHJJM,ANCHORCOOHOOSA}
\begin{equation}
\label{eqn:ch6eq7}
{\sigma_{CC}^{\nu({\overline\nu})N}=\int\limits_0^1 dx \int\limits_0^{xs} dQ^{2} \frac{d^2 \sigma_{CC}^{\nu({\overline\nu})N}}{dxdQ^{2}}}
 \end{equation}

At one loop level (LO), the Callan Gross Relation $F_1=(\frac{1}{2x})F_2$ holds. So obviously, $F_{L}^{CC}=0$ \cite{FIOREJKPPPR12}. 

Hence, putting the value of $Y_{\pm}$ in  eq. (\ref{eqn:ch6eq4}) and considering $t=\log\left(\frac{Q^2}{\Lambda ^2}\right)$ as in Subsection \ref{subsecchap6:Approximate solutions of DGLAP equations:}, we get 
\begin{eqnarray}
\label{eqn:ch6eq8}
\frac{d^2 \sigma_{CC}^{\nu({\overline\nu})N}}{dxdQ^{2}}&=&\frac{G_F^2}{2 \pi}\left(1+\frac{Q^2}{M_W^{2}}\right)^{-2}\frac{1}{x} \left[\left(1-y+\frac{y^2}{2}\right)F_2^{{\nu({\overline\nu})}^{CC}}(x,t)\right.\nonumber\\
& &\left. \pm y(1-\frac{y}{2}) x F_3^{{\nu({\overline\nu})}^{CC}}(x,t)\right]
\end{eqnarray}

Using eq. (\ref{eqn:ch6eq1}) and eq. (\ref{eqn:ch6eq2}), eq. (\ref{eqn:ch6eq8}) can be re-written as,
\begin{eqnarray}
\label{eqn:ch6eq9}
\frac{d^2 \sigma_{CC}^{\nu({\overline\nu})N}}{dxdQ^{2}}&=&\frac{G_F^2}{2 \pi}\left(1+\frac{Q^2}{M_W^{2}}\right)^{-2}\frac{1}{x} \left[\left(1-y+\frac{y^2}{2}\right)F_2^{{\nu({\overline\nu})}^{CC}}(x,t_0)\left(\frac {t}{t_0}\right)^{k(x,t)}   \right.\nonumber\\
& & \left. \times \frac{\left[X^{S}(x)\right]^{k(x,t)}}{\left[X^{S}(x)\right]^{k(x,t_0)}} \pm y(1-\frac{y}{2}) x F_3^{{\nu({\overline\nu})}^{CC}}(x,t_0)\left(\frac {t}{t_0}\right)^{n(x,t)}\right]
\end{eqnarray}

where $k(x,t)$ is identical with ${H^{S}_{expt}(x,t)}$ given by \cite{Campus}
\begin{equation}
\label{eqn:ch5r}
{H^{S}(x,t)=1.284[-4.64 x+ 1.02(1-x)](1.52-1.16s-0.06s^2)}
\end{equation}
and $n(x,t)$ is identical with ${H(x,t)}$ given by \cite{DKCPKD}
\begin{equation}
\label{eqn:ch2eq633}
{H(x,t)=0.380[-5.796 x+ 0.996(1-x)]}
\end{equation}

The $\nu({\overline\nu})N $ structure function $F_i$ can be written as;
\begin{equation}
\label{eqn:ch6eq10}
{F_i=F_i^{light}+F_i^h}
\end{equation}
 
where $F_i^{light}$ provides contributions of all light $(u,d,s)$ quarks and $F_i^h$ refers to contributions of all heavy $(c,b,t)$ quarks. 

We have the following LO expressions in QCD for charged current (CC) interactions \cite{GKR},
\begin{equation}
\label{eqn:ch6eq11}
{F_1^{\nu,light}=\frac{1}{2}\left(\overline{u}+\overline {d}\right)+\frac{1}{2}(d+u)|V_{ud}|^2+s|V_{us}|^2}
\end{equation}
\begin{equation}
\label{eqn:ch6eq12}
{F_2^{\nu,light}=2x{F_1^{\nu,light}}}
\end{equation}
\begin{equation}
\label{eqn:ch6eq13}
{F_3^{\nu,light}=-\left(\overline {u}+\overline {d}\right)+(d+u)|V_{ud}|^2+2s|V_{us}|^2}
\end{equation}

where $u=u(x,Q^2)$ etc. and $V_{ud}$, $V_{us}$ etc. are the relevant CKM matrix elements \cite{PDG,CKMUT,Amser}. We have the following values for the following matrix elements: $V_{ud}= 0.97419 \pm 0.00022$, $V_{us}= 0.2257 \pm 0.0010$. Here we consider the individual contributions from the CC subprocesses of the light quarks i.e. $ W + d \rightarrow u$, $W + s \rightarrow u$, $W + u \rightarrow d$, etc. 
                                  ̄
Let us now consider the heavy quark treatment in a fully massive way \cite{GKR} as we find in case of GRV 98 parton densities \cite{GRVARXIV55} and have the following expressions for heavy quark component. 
\begin{equation}
\label{eqn:ch6eq14}
{F_1^{\nu,heavy}=\frac{1}{2}(d+u)|V_{cd}|^2+s|V_{cs}|^2+c+b(\xi,Q^2+ m_t^2)|V_{tb}|^2}
\end{equation}
\begin{equation}
\label{eqn:ch6eq15}
{F_2^{\nu,heavy}=x(d+u)|V_{cd}|^2+2xs|V_{cs}|^2+2xc+2\xi b(\xi,Q^2+m_t^2)|V_{tb}|^2}
\end{equation}
\begin{equation}
\label{eqn:ch6eq16}
{F_3^{\nu,heavy}=(d+u)|V_{cd}|^2+2s|V_{cs}|^2-2c+2b(\xi,Q^2+m_t^2)|V_{tb}|^2}         
\end{equation}

\begin{figure}[t]
\vspace{0.4in}
\begin{center}
\includegraphics[width=4in]{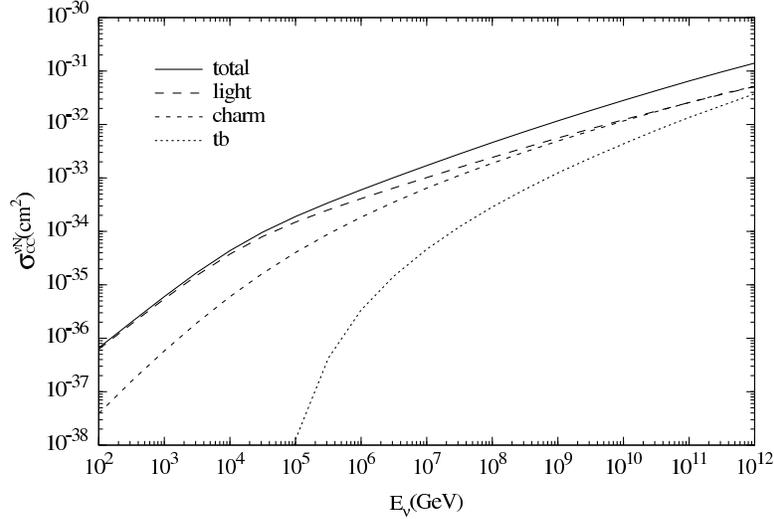}
\end{center}
\vspace{-0.1in}
\caption[Total Neutrino-nucleon interaction CC cross-section versus Energy of the neutrino at NLO along with individual contributions due to light, charm quark transitions as well as contributions due to fully massive tb ]{Total Neutrino-nucleon interaction CC cross-section versus Energy of the neutrino at NLO along with individual contributions due to light, charm quark transitions as well as contributions due to fully massive tb. Figure taken from Ref. \protect\cite{JIMENEZDELREY}} 
\label{fig:russia12}
\end{figure}

Here ${\xi}=\displaystyle{x\left(1+\frac{m_{t}^2}{Q^2}\right)}$, $m_{t}=175$ GeV, $m_{b}=4.5$ GeV, $V_{cd}=0.2256 \pm 0.0010$ \cite{Amser} and contributions like $\displaystyle{\frac{m_c^2}{Q^2}}$  have been neglected. The Charm contribution to the heavy flavour component $F_{i}^{heavy}$ at one loop level takes place through the transition channels $W^{+}s\rightarrow c$ and $W^{+}d\rightarrow c$ and the top bottom contribution comes through the fusion subprocess channel $W^{+}g\rightarrow t\overline{b}$. There are also additional subprocesses $W^{+}g\rightarrow c\overline{s}$ and  $W^{+} + s^{\prime} \rightarrow gc$, etc., where $s_{\nu N}^{\prime} \equiv |V_{cs} |^2 s + |V_{cd} |^2 (d+u)/2$ \cite{JIMENEZDELREY}. Again we neglect the contribution of  $t(x,Q^2)$ since at the relevant $Q^2$ range in the UHE domain, the nucleon contains a very negligible amount of $t\overline{t}$ sea \cite{RCMI111}. It has been found that the contributions from the $c\overline{s}$ sector surpasses the minor contributions provided by heavy $t\overline{b}$ in LO where we take the fully massive $m_{b,t} \not= 0$. This situation prevails for $E_{\nu}\le 10^{8}$ GeV, but when we progressively move towards higher energy and make a switchover from LO to NLO, then we find that LO contribution gets an additional ${\cal{O}}(\alpha_s)$ correction by about $20\%$ due to the NLO massive $t\overline{b}$ contribution \cite{JIMENEZDELREY} as evident in Figure \ref{fig:russia12}.

On the other hand, if we consider heavy quark flavours as massless, as we find in case of GRV 92 parton densities \cite{GRVARXIV33,GRVARXIV44}
\begin{equation}
\label{eqn:ch6eq17}
{F_1^{\nu,heavy}=2c(V_u^2+A_u^2)+2b(V_d^2+A_d^2)}
\end{equation}
\begin{equation}
\label{eqn:ch6eq18}
{F_2^{\nu,heavy}=2xF_1^{\nu,heavy}}
\end{equation}
\begin{equation}
\label{eqn:ch6eq19}
{F_3^{\nu,heavy}=0}
\end{equation}
where $V_u=\frac{1}{2}-\frac{4}{3}\sin^2\theta_W$, $A_u=-A_d=\frac{1}{2}$ and $\sin^2\theta_W=0.232$.

Now considering the minimal contributions provided by the heavy quarks, which is not so prominent as the contributions provided by the light quarks, we only focus on the light quarks in our work.

For $\nu N$ interaction, one uses the relations eq. (\ref{eqn:ch6eq11}) to eq. (\ref{eqn:ch6eq13}) and the following correspondances,
\begin{equation}
\label{eqn:ch6eq20}
{F_{1,2}^{\overline\nu}=F_{1,2}^\nu (q\longleftrightarrow {\overline q})}
\end{equation}
\begin{equation}
\label{eqn:ch6eq21}
{F_{3}^{\overline\nu}=-F_{3}^\nu (q\longleftrightarrow {\overline q})}
\end{equation}

Now putting the explicit parton distributions from eq. (\ref{eqn:ch6eq11}) to eq. (\ref{eqn:ch6eq13}) in eq. (\ref{eqn:ch6eq9}), we have;
\begin{eqnarray}
\label{eqn:ch6eq22}
\frac{d^2 \sigma_{CC}^{\nu(\overline{\nu})N}}{dxdQ^{2}}&=&\frac{G_F^2}{2 \pi}\left(1+\frac{Q^2}{M_W^{2}}\right)^{-2}\frac{1}{x} \left[\left(1-y+\frac{y^2}{2}\right)2x\left\{\frac{1}{2}\left[{\overline u(x,t_0)}+{\overline d(x,t_0)}\right] \right.\right.\nonumber\\
& & \left. +\frac{1}{2}[d(x,t_0)+u(x,t_0)]|V_{ud}|^2+s|V_{us}|^2\right\}\frac{t^{k(x,t)}}{t_0^{k(x,t_0)}}\frac{\left[X^{S}(x)\right]^{k(x,t)}}{\left[X^{S}(x)\right]^{k(x,t_0)}}\nonumber \\
& &  \pm y\left(1-\frac{y}{2}\right)x \left\{-\left[\overline {u}(x,t_0)+\overline {d}(x,t_0)\right]\right.\nonumber \\
& & \left.\left. +[d(x,t_0)+u(x,t_0)]|V_{ud}|^2+2s|V_{us}|^2\right\}\left(\frac {t}{t_0}\right)^{n(x,t)}\right]
\end{eqnarray}

From the kinematics of deep inelastic neutrino scattering, we have
\begin{equation}
\label{eqn:ch6eq23}
{E_{\nu}=\frac{s-M^{2}}{2M}}
\end{equation}
In the UHE region of interest scanned by us, we can reasonably re-write eq. (\ref{eqn:ch6eq23}) as \cite{FIOREJKPPPR12,FIOREJKPPPR1}
\begin{equation}
\label{eqn:ch6eq24}
{E_{\nu}=\frac{s-M^{2}}{2M} \simeq \frac{s}{2M}}
\end{equation}

Again from the kinematics of deep inelastic neutrino scattering, we can reasonably approximate \cite{FIOREJKPPPR12,FIOREJKPPPR1}
\begin{equation}
\label{eqn:ch6eq25}
{y=\frac{Q^2}{x(s-M^{2})} \simeq \frac{Q^2}{xs}}
\end{equation}

By setting proper lower cuts for $Q^{2}$ and $x$ integration \cite{FIOREJKPPPR12,FIOREJKPPPR1} in eq. (\ref{eqn:ch6eq7}), we get 

\begin{equation}
\label{eqn:ch6eq26}
{\sigma_{CC}^{\nu(\overline{\nu})N}=\int\limits_{\frac{Q^2}{s}}^1 dx \int\limits_{Q_0^2}^{s} dQ^{2} \frac{d^2 \sigma_{CC}^{\nu(\overline{\nu})N}}{dxdQ^{2}}}
 \end{equation}

The contribution from $xF_{3}$ at Ultra-low Bjorken $x$ in UHE domain is very small. But still retaining this factor, we have from eqs. (\ref{eqn:ch6eq22})  and (\ref{eqn:ch6eq26}),
\begin{eqnarray}
\label{eqn:ch6eq27}
 \sigma_{CC}^{\nu(\overline{\nu})N}&=&\frac{G_F^2}{2 \pi}\int\limits_{Q_0^2}^{s}dQ^{2}\left(1+\frac{Q^2}{M_W^{2}}\right)^{-2}\int\limits_{\frac{Q^2}{s}}^1\frac{dx}{x} \left[\left(1-\frac{Q^2}{xs}+\frac{Q^4}{2x^{2}s^{2}}\right)\right.\nonumber\\
& & \times 2x\left\{\frac{1}{2}\left[{\overline u(x,t_0)}+{\overline d(x,t_0)}\right]+\frac{1}{2}[d(x,t_0)+u(x,t_0)]|V_{ud}|^2+s|V_{us}|^2\right\} \nonumber \\
& &  \times \frac{t^{k(x,t)}}{t_0^{k(x,t_0)}}\frac{\left[X^{S}(x)\right]^{k(x,t)}}{\left[X^{S}(x)\right]^{k(x,t_0)}}\pm \frac{Q^2}{xs}(1-\frac{Q^2}{2xs})x\left\{-\left[\overline {u}(x,t_0)+\overline {d}(x,t_0)\right]\right. \nonumber \\
& &\left.\left. +[d(x,t_0)+u(x,t_0)]|V_{ud}|^2+2s|V_{us}|^2\right\}\left(\frac {t}{t_0}\right)^{n(x,t)}\right]
\end{eqnarray}
Here $u(x,t_0)$, $d(x,t_0)$ etc. represent parton distributions at various values of $x$ and input value $t_0$.
\subsubsection{Charged Current-Numerical Solution (exact):}

Putting the explicit parton distributions from eq. (\ref{eqn:ch6eq11}) to eq. (\ref{eqn:ch6eq13}) in eq. (\ref{eqn:ch6eq8}), we get the expression of double differential cross-section as
\begin{eqnarray}
\label{eqn:ch6eq27AA}
\frac{d^2 \sigma_{CC}^{\nu(\overline{\nu})N}}{dxdQ^{2}}&=&\frac{G_F^2}{2 \pi}\left(1+\frac{Q^2}{M_W^{2}}\right)^{-2}\frac{1}{x} \left[\left(1-y+\frac{y^2}{2}\right)2x\left\{\frac{1}{2}\left[\overline {u}(x,t)+\overline {d}(x,t)]\right) \right.\right.\nonumber\\
& & \left. +\frac{1}{2}[d(x,t)+u(x,t)]|V_{ud}|^2+s|V_{us}|^2\right\}\pm y\left(1-\frac{y}{2}\right)x \nonumber \\
& & \left. \left\{-\left[\overline {u}(x,t)+\overline {d}(x,t)\right]+[d(x,t)+u(x,t)]|V_{ud}|^2+2s|V_{us}|^2\right\}\right]\nonumber \\
\end{eqnarray}

Now putting the expression for double differential cross-section from eq. (\ref{eqn:ch6eq27AA}) in eq. (\ref{eqn:ch6eq26}), we get

\begin{eqnarray}
\label{eqn:ch6eq27BB}
 \sigma_{CC}^{\nu(\overline{\nu})N}&=&\frac{G_F^2}{2 \pi}\int\limits_{Q_0^2}^{s}dQ^{2}\left(1+\frac{Q^2}{M_W^{2}}\right)^{-2}\int\limits_{\frac{Q^2}{s}}^1\frac{dx}{x} \left[\left(1-\frac{Q^2}{xs}+\frac{Q^4}{2x^{2}s^{2}}\right)2x\left\{\frac{1}{2}\left[{\overline u(x,t)} \right.\right.\right.\nonumber\\
& & \left.\left.+{\overline d(x,t)}\right]+\frac{1}{2}[d(x,t)+u(x,t)]|V_{ud}|^2+s|V_{us}|^2\right\}\pm \frac{Q^2}{xs}\left(1-\frac{Q^2}{2xs}\right)x \nonumber \\
& & \left. \left\{-\left[\overline {u}(x,t)+\overline {d}(x,t)\right]+[d(x,t)+u(x,t)]|V_{ud}|^2+2s|V_{us}|^2\right\}\right]
\end{eqnarray}
Here $u(x,t)$, $d(x,t)$ etc. represent parton distributions at various values of $x$ and  $t$.

\subsubsection{ Neutral current-Analytical Solution:}
\label{subsubsecchap6:Neutral Current}

The total cross-section for UHE NC interaction $(\sigma_{NC})$ in an isoscalar target $N [\equiv\frac {(p+n)}{2}]$ is similar to eq. (\ref{eqn:ch6eq26}), viz

\begin{equation}
\label{eqn:ch6eq28}
{\sigma_{NC}^{\nu({\overline\nu})N}=\int\limits_{\frac{Q^2}{s}}^1 dx \int\limits_{Q_0^2}^{s} dQ^{2} \frac{d^2 \sigma_{NC}^{\nu({\overline\nu})N}}{dxdQ^{2}}}
 \end{equation}

where

\begin{eqnarray}
\label{eqn:ch6eq30}
\frac{d^2 \sigma_{NC}^{\nu({\overline\nu})N}}{dxdQ^{2}}&=&\frac{G_F^2}{2 \pi}\left(1+\frac{Q^2}{M_Z^{2}}\right)^{-2}\frac{1}{x} \left[\left(1-y+\frac{y^2}{2}\right)F_2^{{\nu({\overline\nu})}^{NC}}(x,t_0)\left(\frac {t}{t_0}\right)^{k(x,t)}\right.\nonumber\\
& &\left. \times \frac{\left[X^{S}(x)\right]^{k(x,t)}}{\left[X^{S}(x)\right]^{k(x,t_0)}} \pm y(1-\frac{y}{2}) x F_3^{{\nu({\overline\nu})}^{NC}}(x,t_0)\left(\frac {t}{t_0}\right)^{n(x,t)}\right]
\end{eqnarray}
to be compared with eq. (\ref{eqn:ch6eq9}) for charged current.

We have the following LO expressions for structure functions involving Neutral currents (NC) \cite{GKR}:

\begin{equation}
\label{eqn:ch6eq31}
{2F_1^{\nu,light}=\frac{1}{2}(u+\overline{u}+d+\overline {d})(V_u^2+A_u^2)+\frac{1}{2}(u+\overline{u}+d+\overline {d}+4s)(V_d^2+A_d^2)}
\end{equation}
\begin{equation}
\label{eqn:ch6eq32}
{F_2^{\nu,light}=2x{F_1^{\nu,light}}}
\end{equation}
\begin{equation}
\label{eqn:ch6eq33}
{2F_3^{\nu,light}=2(u_v+d_v)(V_u A_u+V_d A_d)}
\end{equation}
where the expressions for $V_u$ and $V_d$ as well as the values of $A_u$ and $A_d$ have been mentioned beforehand. 

As in the case of charged current, we focus only on the contributions from light quarks.

Putting the values of $V_u$, $A_{u}$, $V_{d}$ and $A_d$ in eqs. (\ref{eqn:ch6eq31}), (\ref{eqn:ch6eq32}) and (\ref{eqn:ch6eq33}) and using eq. (\ref{eqn:ch6eq30}) and eq. (\ref{eqn:ch6eq28}), we finally get;

\begin{eqnarray}
\label{eqn:ch6eq34}
 \sigma_{NC}^{\nu(\overline{\nu})N}&=&\frac{G_F^2}{2 \pi}\int\limits_{Q_0^2}^{s}dQ^{2}\left(1+\frac{Q^2}{M_Z^{2}}\right)^{-2}\int\limits_{\frac{Q^2}{s}}^1\frac{dx}{x} \left[\left(1-\frac{Q^2}{xs}+\frac{Q^4}{2x^{2}s^{2}}\right) \left\{0.1432 x \right.\right.\nonumber \\
& & \times \left[u(x,t_0)+\overline {u}(x,t_0)+d(x,t_0)+\overline{d}(x,t_0)\right]+0.1849 x \nonumber\\
& & \left.\times \left[u(x,t_0)+\overline{u}(x,t_0)+d(x,t_0)+\overline{d}(x,t_0)+4s\right]\right\}\frac{t^{k(x,t)}}{t_0^{k(x,t_0)}}\frac{\left[X^{S}(x)\right]^{k(x,t)}}{\left[X^{S}(x)\right]^{k(x,t_0)}} \nonumber \\ 
& &  \pm \frac{Q^2}{xs}\left(1-\frac{Q^2}{2xs}\right)
\left\{0.268 x \left[u(x,t_0)-\overline {u}(x,t_0)+d(x,t_0)-\overline{d}(x,t_0)\right]\right\}\nonumber \\
& & \left.\times \left(\frac {t}{t_0}\right)^{n(x,t)}\right]
\end{eqnarray}

which can be re-expressed as

\begin{eqnarray}
\label{eqn:ch6eq35}
 \sigma_{NC}^{\nu(\overline{\nu})N}&=&\frac{G_F^2}{2 \pi}\int\limits_{Q_0^2}^{s}dQ^{2}\left(1+\frac{Q^2}{M_Z^{2}}\right)^{-2}\int\limits_{\frac{Q^2}{s}}^1\frac{dx}{x} \left[\left(1-\frac{Q^2}{xs}+\frac{Q^4}{2x^{2}s^{2}}\right) \left\{0.1432 x  \right.\right.\nonumber\\
& & \times \displaystyle{ \left[\left(u_v(x,t_0)+d_v(x,t_0)\right)+2  \left(\overline{u}(x,t_0)+\overline{v}(x,t_0)\right)\right]}+0.1849 x \nonumber\\
& &\left.\times\left[\left(u_v(x,t_0)+d_v(x,t_0)\right)+2 \left(\overline{u}(x,t_0)+\overline{v}(x,t_0)+4s\right)\right]\right\}\frac{t^{k(x,t)}}{t_0^{k(x,t_0)}} \nonumber \\
 & & \left. \times \frac{\left[X^{S}(x)\right]^{k(x,t)}}{\left[X^{S}(x)\right]^{k(x,t_0)}}\pm \frac{Q^2}{xs}\left(1-\frac{Q^2}{2xs}\right) 
 \left\{0.268 x \left[u_v(x,t_0)+d_v(x,t_0)\right]\right\}\left(\frac {t}{t_0}\right)^{n(x,t)}\right]\nonumber \\
\end{eqnarray}

\subsubsection{Neutral Current-Numerical Solution (exact):}
We have in case of neutral current the following equation
\begin{eqnarray}
\label{eqn:ch6eq35AAAA}
\frac{d^2 \sigma_{NC}^{\nu({\overline\nu})N}}{dxdQ^{2}}&=&\frac{G_F^2}{2 \pi}\left(1+\frac{Q^2}{M_Z^{2}}\right)^{-2}\frac{1}{x} \left[\left(1-y+\frac{y^2}{2}\right)F_2^{{\nu({\overline\nu})}^{NC}}(x,t)\right.\nonumber\\
& &\left. \pm y\left(1-\frac{y}{2}\right) x F_3^{{\nu({\overline\nu})}^{NC}}(x,t)\right]
\end{eqnarray}
similar to eq. (\ref{eqn:ch6eq8}) for charged current.

Now putting the explicit parton distributions from eq. (\ref{eqn:ch6eq31}) to eq. (\ref{eqn:ch6eq33}) in eq. (\ref{eqn:ch6eq35AAAA}) and then using eq. (\ref{eqn:ch6eq28}), we get the expression for neutral current cross-section as

\begin{eqnarray}
\label{eqn:ch6eq35Times}
 \sigma_{NC}^{\nu(\overline{\nu})N}&=&\frac{G_F^2}{2 \pi}\int\limits_{Q_0^2}^{s}dQ^{2}\left(1+\frac{Q^2}{M_Z^{2}}\right)^{-2}\int\limits_{\frac{Q^2}{s}}^1\frac{dx}{x} \left[\left(1-\frac{Q^2}{xs}+\frac{Q^4}{2x^{2}s^{2}}\right) \left\{0.1432 x \right.\right.\nonumber \\
& & \times \left[u(x,t)+\overline {u}(x,t)+d(x,t)+\overline{d}(x,t)\right]+0.1849 x \nonumber\\
& & \left.\times \left[u(x,t)+\overline{u}(x,t)+d(x,t)+\overline{d}(x,t)+4s\right]\right\} \nonumber \\ 
& & \left. \pm \frac{Q^2}{xs}\left(1-\frac{Q^2}{2xs}\right)
\left\{0.268 x \left[u(x,t)-\overline {u}(x,t)+d(x,t)-\overline{d}(x,t)\right]\right\}\right]\nonumber \\ 
\end{eqnarray}

which can be re-expressed as

\begin{eqnarray}
\label{eqn:ch6eq35ROMAN}
 \sigma_{NC}^{\nu(\overline{\nu})N}&=&\frac{G_F^2}{2 \pi}\int\limits_{Q_0^2}^{s}dQ^{2}\left(1+\frac{Q^2}{M_Z^{2}}\right)^{-2}\int\limits_{\frac{Q^2}{s}}^1\frac{dx}{x} \left[\left(1-\frac{Q^2}{xs}+\frac{Q^4}{2x^{2}s^{2}}\right) \left\{0.1432 x  \right.\right.\nonumber\\
& & \times\displaystyle{ \left[\left(u_v(x,t)+d_v(x,t)\right)+2  \left(\overline{u}(x,t)+\overline{v}(x,t)\right)\right]}+0.1849 x\nonumber\\
& &\left.\times\left[\left(u_v(x,t)+d_v(x,t)\right)+2  \left(\overline{u}(x,t)+\overline{v}(x,t)+4s\right)\right]\right\} \nonumber \\
 & & \left. \pm \frac{Q^2}{xs}\left(1-\frac{Q^2}{2xs}\right) 
 \left\{0.268 x \left[u_v(x,t)+d_v(x,t)\right]\right\}\right]
\end{eqnarray}

\subsection{Role of the gauge-boson propagator:}
\label{subsecchap6:Role of the gauge-boson propagator:}
For neutrino energy $\displaystyle{E_{\nu}\ll \frac{M_W^2}{2M}}$ (or say $Q^{2}\ll M_W^{2}$), it is found that $\sigma \propto E_{\nu}$. This is because of the fact that in the linear regime at low $Q^2$, the gauge boson propagator and the PDFs do not almost depend on $Q^{2}$ and then for $E_{\nu}\ge 10^{5}$ GeV,  $\sigma$ is found to be approximately proportional to $E_{\nu}^{0.4}$. When $E_{\nu}$ is greater than $10^{3}$ GeV i.e. for $\displaystyle{E_{\nu}\gg \frac{M_W^2}{2M}}$ (or say $Q^{2}\gg M_W^{2}$), the gauge boson propagator decreases sharply making the double differential cross-section to die off fast \cite{RCMI1}. Consequently the gauge boson propagator puts a restriction on $Q^{2}$ to assume values near $M_{W}^{2}$ and thus $Q^2$ integration region  shrinks, thus making the upper limit of the $Q^2$ integral  proportional to $M_W^{2}$ \cite{FIOREJKPPPR12,MRISGSMSWV}. Thus at ultra high energy, $Q^2$ dependence of the gauge boson propagator and the PDFs becomes prominent. In one hand, there is a power suppression from the boson propagator, while on the other hand there is a logarithmic growth of the PDFs. In overall, the propagator dominates and it leads to the generation of the dampness of the total cross-section \cite{MHRENO12}. It is pertinent to note that the effective interval in the fractional parton momentum $x$ gets limited to the region around $\displaystyle{\frac{M_W^{2}}{2 M E_{\nu}}}$. It is interesting to note that in the $(x, Q^2)$ phase space at the high energy of the tone of $E_{\nu}\ge 10^{5}$ GeV, the cross section gets dominated by the behavior of the quark densities at $\displaystyle{x=\frac{M_W^{2}}{2 M E_{\nu}}}$ and $Q^{2}\sim M_{W}^{2}$ \cite{RCMI}. For the neutrino energy $E_{\nu} \sim 10^{12}$ GeV, we have $x \sim 10^{-8}$.

Though a cut off in $Q^{2}$ is automatically created at $Q^{2}\sim M_{W}^{2}$ due to the propagator effect, but we have carried out the integration upto the upper limit $s=2ME_{\nu}$ (eq. (\ref{eqn:ch6eq24})) to take into consideration the unaccounted part of the results of cross-section for $Q^{2}> M_{W}^{2}$. Of course, we have once checked that the difference is not too much, rather minimal by putting the upper limit at $Q^{2}\sim M_{W}^{2}\sim 10^{4}$ GeV$^2$, thus justifying the argument of the validity of the taking the upper limit of the $Q$-integration around $Q^{2}\sim M_W^{2}$ at ultra high energy of the neutrino (as well as antineutrino).

\subsection{Gluon recombination, saturation and unitarity:}
\label{subsecchap6: gluon recombination,saturation and uniterity}
   It is interesting to note that in these approaches, cross-section increases sharply with $E_{\nu}$ in a fast power-law-like growth at the highest energies. But this rapid growth gets slowed down when we consider the upper extreme end of the low $x$ region and  unitarization of the cross section takes place \cite{RAHULBASU,EMHJJM,MRISGSMSWV}. At extremely low values of $x$, the  density of parton (gluon) phase space will increase. Consequently, there will be overlapping in transverse space and $gg \sim g$ recombination processes will trigger up. The characterizing scale of the saturation region (regime in $x$ and $Q^{2}$) is given by,

\begin{figure}
\vspace{-0.40in}
\begin{center}
\includegraphics[width=4.5in]{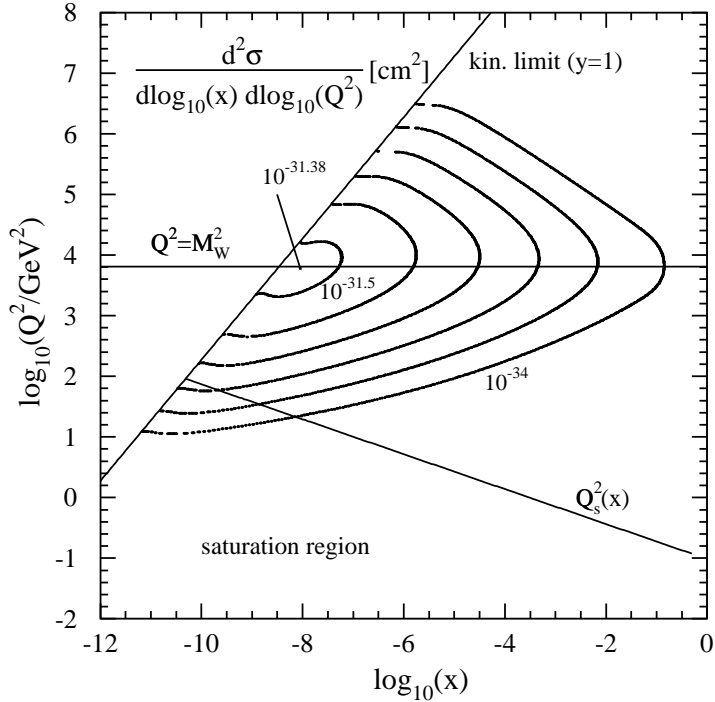}\\
\end{center}
\vspace{-0.15in}
\caption[Contour plot of the neutrino-nucleon cross section in the ($x, Q^2$) plane.]{The figure in the above is the contour plot of the neutrino-nucleon cross section $\left(\frac{d^2 \sigma}{dQ^{2}dx}\right)$  in the ($x, Q^2$) plane. Saturation region is plotted from Eq. (\ref{eqn:cheq19}). Figure taken from Ref. \protect\cite{MRISGSMSWV}}. 
\label{fig:chapt1contourplot}
\end{figure}

\begin{equation}
\label{eqn:cheq19}
{Q_{s}^2(x)=(1 GeV)^2\left(\frac{x_0}{x}\right)^{\lambda}}
\end{equation}

The value of the parameters $\lambda$ and $x_0$ are obtained as $0.288$ and $3.04 \times 10^{-4}$ respectively by a fit to HERA data. The domain of saturation starts with those values of $Q$ which are lower than the saturation scale $Q_s(x)$ at a given $x$. It is clear from the above equation that with the increasing/decreasing values of $x$, saturation takes place at smaller/larger values of $Q^{2}$. In the highest sensitive UHE neutrino interaction domain $x\simeq 10^{-8}$, saturation scale is around $Q_s^2=20$ $GeV^{2}$. It is evident from the Figure \ref{fig:chapt1contourplot} that total cross-section is dominated by the scale $Q\sim M_{W,Z}$. An obvious question is that what is the effect of gluon recombination on the total cross-section? Interestingly, the contribution of the saturation region to the total cross-section is not so much appreciable \cite{EMHJJM,MRISGSMSWV}. Even at very high energy say $E_{\nu}=10^{12}$\, GeV, the contribution is much lower than $1\%$. The region  of $Q^2 < 1 \,\,GeV^{2}$ and consequently the saturation region at $E=10^{12}$ GeV contribute very little to the total cross-section. We must realize that at the time of evaluation of total cross-section, we carry out our integration over large range of values of $x$ and $Q^{2}$ in the ($x,Q^{2}$) phase space and since the saturation region comprises of a very small section of the phase space, hence the effect is marginal. But the importance of saturation lies in retarding the growth of cross-section so that it does not cross the unitarity bound (Froissart bound).  This is quite obvious from the Figure \ref {fig:chapt1contourplot}.                      

\subsection{Small $x$ extrapolation}
\label{subsecchap6: Small $x$ extrapolation}
Since the UHE neutrinos coming from various sources have in general energy more than $10^5$ GeV, so perturbative calculations should be carried out using parton distribution functions at Ultra-low Bjorken $x$ ($x < 10^{-5})$. Particularly, the sensitivity of UHE neutrino interactions are greatest in the domain $x \sim 10^{-8}$ and $Q^2 \sim M_{W}^{2}$. But high energy measurements of deep inelastic scattering (DIS) of lepton-nucleon at the collider HERA \cite{H1ABT,H1AID1,H1AHMED,H1AID2,H1ADLOFF1,H1ADLOFF2,H1ADLOFF3,H1ADLOFF4,H1ADLOFF5} at DESY have provided interesting data from time to time. To the best of its ability, it has provided data for $x\ge 10^{-5}$. The greatest available energy  $\sqrt{s_{ep}}$ available at HERA is $319 \,$ GeV, which is far below the expected UHE neutrino-nucleon collision energies of about $\sqrt{s_{\nu N}}=10^{^6} $ GeV \cite{GKR,FIOREJKPPPR}. So it is not possible to scan the ultra-high energy region and evaluate the UHE cross-section from the present DIS experiments alone.
                                          
   There are two different approaches to determine the cross-sections for UHE neutrino-nucleon interaction. One way is to parametrize the parton distribution functions by global fitting to a rich set of available data at low values of $Q^2$ (say $Q^2=Q_0^2)$. The numerical solutions of the DGLAP equations \cite {GRIBOVLIPA,LIPATOV,DOKSH,AP,GALTA} are then used to evolve parton densities at higher values of $Q^2$ ($Q^2>Q_0^2)$. This $Q^2$-evolution of small-$x$ parton distribution is not the right choice to scan the region of UHE neutrino-nucleon interactions as the evaluation is valid in the region of $x>10^{-5}$. Extensive extrapolation \cite{RCMI1,GKR,RAMCFR,RCMI111,RCMI,RCMI11,CACRPAZA,RAMCFR0} is then carried out to determine the parton densities from the ($x$, $Q^{2}$ ) domain of terrestrial accelerator HERA to the ultrasmall values of Bjorken $x$, where there is no data. But unfortunately, this type of extrapolation brings about some uncertainties in the estimated UHE neutrino-nucleon cross-sections \cite{GKR} measured with the help of neutrino telescopes.  This is because we are not so sure about what is happening in the region $x\rightarrow 0$ and our ignorance forces us to take different assumtions ( or educated guesses) about the behaviour of the distribution functions at ultra-small $x$, leading to the large variations of different evaluated cross-sections.

As an example of the first approach, we can mention the case of Gandhi et al \cite{RCMI1,RCMI111,RCMI,RCMI11} who used CTEQ3 and CTEQ4 parton distributions \cite{LAIHL11,LAIHL1122} and MRS parton distrbutions \cite{ADMWJSTRGRO,ADMWJSTRGRO1,ADMWJSTRGRO12,ADMWJSTRGRO123} ( MRS $A^{,}, G , D_{-,}$ and $D_{-}^{,}$) to determine the charged current cross-section $\sigma_{CC}$ and the neutral current cross-section $\sigma_{NC}$. On the other hand, in the second approach, Gluck, Kretzer and Reya  \cite{GKR} applied QCD-inspired radiative parton model \cite{GRVARXIV55A,GRVARXIV55,GRVARXIV33,GRVARXIV44,GRVARXIV22,GRV98} to evaluate parton densities directly for $x \le 10^{-5}$. The model utilizes the principle of QCD dynamics to extrapolate parton densities to kinematical regions as yet unexplored by the present day terrestrial experiment and to go for respective fits. Parton densities obtained within the framework of this model only are claimed to be free from the ambiguity of the earlier approach. Here one calculates the parton densities in the region $x\rightarrow 0$ by the direct application of QCD dynamics, without introducing any fit parameter in that region. It may be mentioned that though MRS and CTEQ parametrisations do not work at all for values of $x$ below $10^{-5}$, but GRV parametrisation works well for values of $x$ upto $10^{-9}$ \cite{RAHULBASU}. 

At Ultra-low Bjorken $x$, sea quarks will hugely multiply in a dominating way and these arise from gluon splitting into quark-antiquark pairs. So the technique of gluon extrapolation will provide one a direction to carry out sea quark extrapolation. If the gluon extrapolation is carried out at some input value $Q_0^2$ in the following way \cite{ELLISKUNLEVINMEZ},
\begin{equation}
xg(x,Q_0^2)\sim A(Q_0^2)\, x^{-\lambda}\ \quad\quad x \ll 1\ ,
\end{equation}
then at higher values of $Q^2$, we can extrapolate gluons similarly
\begin{equation}
xg(x,Q^2)\sim A(Q^2)\, x^{-\lambda}\ .
\end{equation}
We can extrapolate sea quarks in a similar approach \cite{MHRENO12111}
\begin{equation}
x{q}(x,Q^2)=\Biggl( \frac{x_{min}}{x}\Biggr)^{\lambda}
x{q}(x_{min},Q^2)\ .
\end{equation} 
One can determine $\lambda$ from the PDFs at $Q^2=M_W^2$ for each flavor.
It is pertinent to know that we have developed our formalism in this paper in leading order only, so we will use LO parton densities in this case. In case of NLO, one has to use NLO parton densities strictly, with the appropriate addition of convolutions with the fermionic Wilson coefficient
$C_q$ and the NLO contribution $C_g \otimes g$ \cite{FURPETRONZIO1}, though these additional contributions never provide more than 2\% contributions \cite{GKR}.
\begin{figure}[t]
\begin{center}
\includegraphics[width=4.5in]{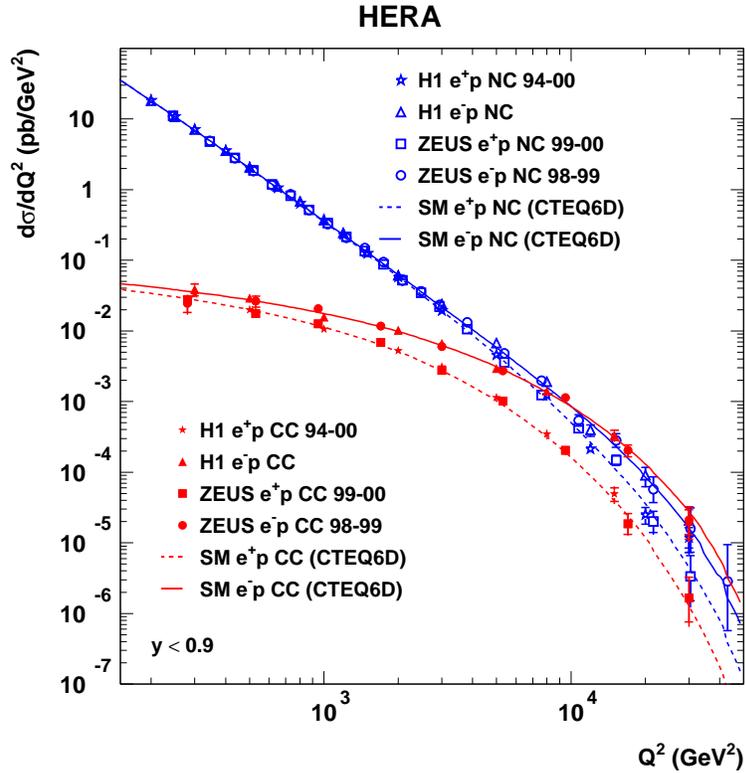}
\end{center}
\vspace{-0.1in}
\caption[Neutral and charged current cross sections as measured by H1 and ZEUS alongwith Standard Model predictions]{Neutral and charged current cross sections as measured by H1 and ZEUS alongwith Standard Model predictions.  Figure taken from Ref. \protect\cite{SASCAR}} 
\label{fig:Raman}
\end{figure} 
As a simplified case, we have not gone for small $x$ extrapolation in this paper with the assumption that there is not too much remarkable difference between the extrapolated and the non-extrapolated results.

\begin{figure}
\begin{center}
\includegraphics[width=4.7in]{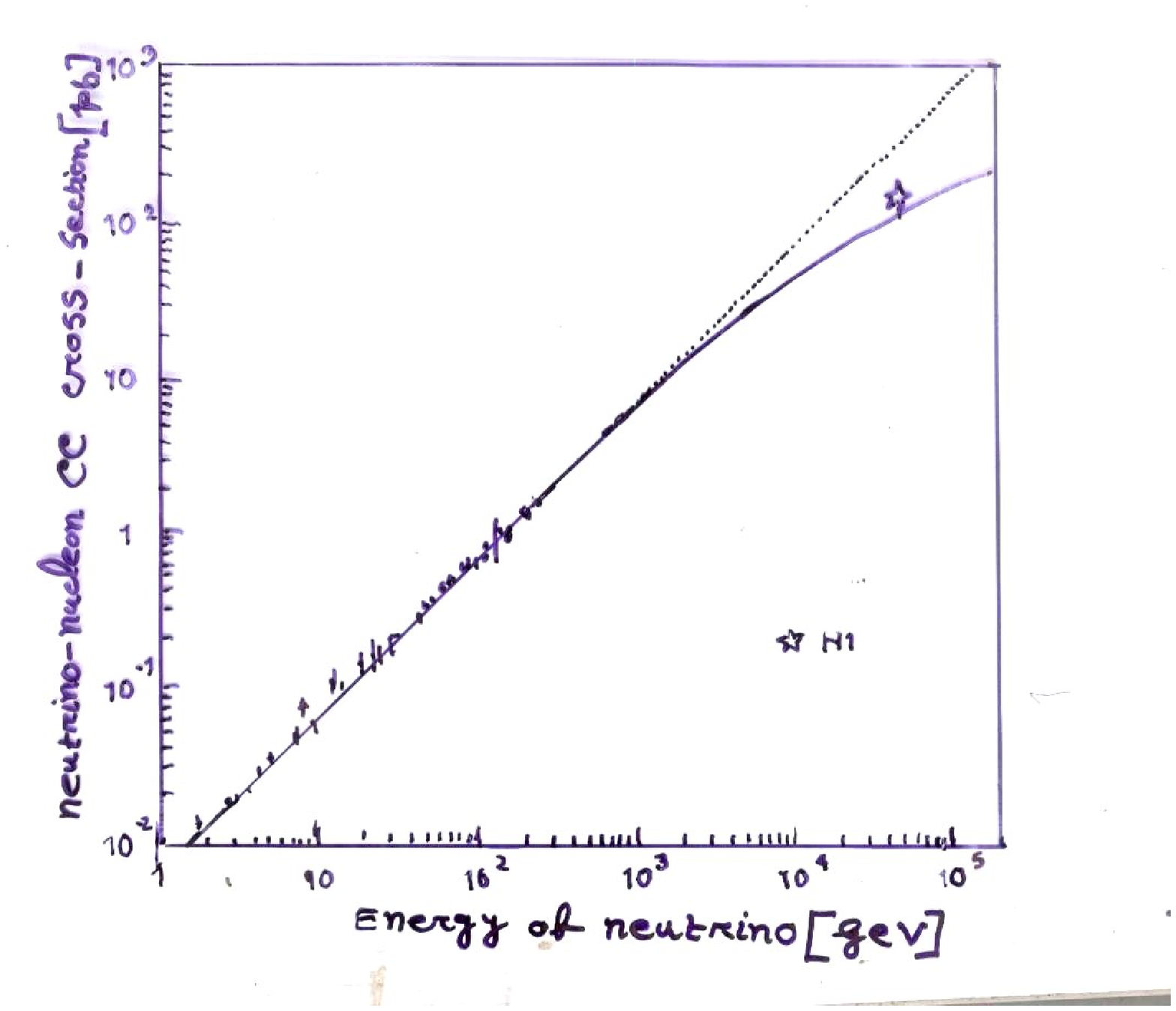}
\end{center}
\vspace{-0.1in}
\caption[Equivalent charged current cross sections in deep inelastic neutrino-nucleon scattering as measured by H1.]{Equivalent charged current cross sections in deep inelastic neutrino-nucleon scattering as measured by H1. Figure taken from Ref. \protect\cite{HAST}. Similar figure is also obtained in \protect\cite{MAHU}} 
\label{fig:meghnad}
\end{figure} 
\subsection{Equivalent neutrino-nucleon cross-section}        
\label{subsecchap6: Equivalent neutrino-nucleon cross-section}
The two detector collaborations H1 and ZEUS at HERA at DESY in Hamburg recorded integrated luminosity of 1 $pb^{-1}$ from May to October, 1993. The amount was twenty times the luminosity that was recorded in 1992 during its run for the first time, where the energy of the electron beam was 26.7 GeV \cite{HAST}. In 1993,  energy of the electrons was raised to 27.5 GeV, which were allowed to collide with 820 GeV protons, where the centre of mass energy ($\sqrt{s}$) was around 300 GeV. From July 18, 1994, electron beam was replaced by positron beam due to the limitation imposed by the lifetime of the electron beam \cite{MAHU}. In the HERA II phase in 2000-2001, the collider as well as experiments were upgraded and consequently 920 GeV proton beam was allowed to collide with 27.6 GeV electron/positron beam with the centre-of-mass energy 319 GeV and integrated ‎‎luminosity around 150 $pb^{-1}$ \cite{SASCAR}. Recent results from HERA shows H1 luminosity as 181 $pb^{-1}$ in $e^{-}p$ and 294 $pb^{-1}$ in $e^{+}p$ scatterings respectively \cite{KKGER}.

On the basis of data collected by H1 and ZEUS, the neutral and charged current cross-sections were calculated during the HERA I phase (1994-2000) and was found to coincide nicely with the electroweak Standard Model predictions based on CTEQ6D parton density functions \cite{SASCAR} as revealed in Figure \ref{fig:Raman}.

From neutrino-nucleon scattering experiments in the eighties, weak charged currents were studied, where it was found that the cross-section is directly proportional to neutrino energy \cite{DHPH}. This linear behaviour of cross-section clearly pinpoints the fact that the sensivity of the neutrino energy to the $W$ Propagator effect is extremely small and hence negligible. The field of electroweak physics was opened for the first time when charged current reaction
$ep \rightarrow \nu X$ was studied at HERA, where the role of $W$ Propagator became important becauge of the high momentum transfer comparable to $Q^2=M_W^2= (80 \,GeV)^2$ and the equivalent fixed target energy of the HERA  collider was 50 TeV. Since this process is just the inverse of neutrino-nucleon scattering, the measured $ep$ cross-section could easily be converted into an equivalent $\nu N$ cross-section at 50 TeV \cite{HAST,MAHU}. Interestingly, this value of equivalent cross-section comes very close to the value of cross-section as predicted by Gandhi et al in the recent past. In Figure \ref{fig:meghnad}, low energy neutrino data has been represented by the crosses, whereas the full and open stars represents the equivalent $\nu N$ cross-section as obtained from the HERA experiment. The extrapolation from low energies has produced the straight line at higher and higher energies assuming $M_W=\infty$, but when the effect of $W$ propagator is considered, one gets the bent curve which represents the predicted cross-section. The $\nu N$ cross-section corresponding to equivalent fixed target energy of 50 TeV is represented by H1 point in the Figure \ref{fig:meghnad}.

\section{ Results and discussions:}
\label{secchap6:Results and discussions}

To estimate the cross-sections for charged and neutral current from eqs. (\ref{eqn:ch6eq26}) and (\ref{eqn:ch6eq28}), we use MRST 2004 f4 LO input parton distributions at $Q^2 = \mu_{LO}^2 = 1\,\, GeV^{2}$ \cite{MRST2222,MRSTDURHAM} 

Putting these distributions in eqs. (\ref{eqn:ch6eq27}) and (\ref{eqn:ch6eq35}) and using eqs. (\ref{eqn:ch6eq26}) and (\ref{eqn:ch6eq28}) to integrate over $x$ and $Q^2$, we obtain the values of cross-sections for $\nu N$ (as recorded in Table \ref{table:ch6tab1}) and $\overline{\nu} N$ (as recorded in Table \ref{table:ch6tab2})  interactions both for charged current and neutral current respectively .

As expected, for $10 \le E_{\nu} \le 10^{3}$ \,GeV, the cross-section rises linearly with energy as discussed in Subsection \ref{subsecchap6:Role of the gauge-boson propagator:} and \ref{subsecchap6: Equivalent neutrino-nucleon cross-section}.

Then the cross-section versus energy relation have the following power law forms:

\begin{eqnarray}
\label{eqn:ch6eq44analytical}
{\sigma_{CC}^{\nu N}=\left\{{\begin{array}{r@{\quad:\quad}l}1.70 \times 10^{-35}\,\,cm^{2}\displaystyle{\left(\frac{E_{\nu}}{1\,\,GeV}\right)^{0.199}} &  10^{5}\le E_{\nu} \le 10^{8} \,GeV \\  2.24 \times 10^{-35}\,\,cm^{2}\displaystyle{\left(\frac{E_{\nu}}{1\,GeV}\right)^{0.183}} &  10^{8}\le E_{\nu} \le 10^{12} \,GeV \end{array}}\right.}
\end{eqnarray}

\begin{eqnarray}
\label{eqn:ch6eq45analytical}
{\sigma_{NC}^{\nu N}=\left\{{\begin{array}{r@{\quad:\quad}l}3.18\times 10^{-36}\,\,cm^{2}\displaystyle{\left(\frac{E_{\nu}}{1\,\,GeV}\right)^{0.213}} &  10^{5}\le E_{\nu} \le 10^{8} \,GeV \\  5.28 \times 10^{-36}\,\,cm^{2}\displaystyle{\left(\frac{E_{\nu}}{1\,GeV}\right)^{0.185}} &  10^{8}\le E_{\nu} \le 10^{12} \,GeV \end{array}} \right.}
\end{eqnarray}

\begin{eqnarray}
\label{eqn:ch6eq46analytical}
{\sigma_{CC}^{\overline{\nu}N}=\left\{{\begin{array}{r@{\quad:\quad}l}1.02\times 10^{-35}\,\,cm^{2}\displaystyle{\left(\frac{E_{\overline{\nu}}}{1\,\,GeV}\right)^{0.231}} &  10^{5}\le E_{\overline{\nu}} \le 10^{8} \,GeV \\  2.23 \times 10^{-35}\,\,cm^{2}\displaystyle{\left(\frac{E_{\overline{\nu}}}{1\,GeV}\right)^{0.183}} &  10^{8}\le E_{\overline{\nu}} \le 10^{12} \,GeV \end{array}} \right.}
\end{eqnarray}

\begin{eqnarray}
\label{eqn:ch6eq47analytical}
{\sigma_{NC}^{\overline{\nu}N}=\left\{{\begin{array}{r@{\quad:\quad}l}1.97 \times 10^{-36}\,\,cm^{2}\displaystyle{\left(\frac{E_{\overline{\nu}}}{1\,\,GeV}\right)^{0.239}} &  10^{5}\le E_{\overline{\nu}} \le 10^{8} \,GeV \\  5.28 \times 10^{-36}\,\,cm^{2}\displaystyle{\left(\frac{E_{\overline{\nu}}}{1\,GeV}\right)^{0.185}} &  10^{8}\le E_{\overline{\nu}} \le 10^{12} \,GeV \end{array}} \right.}
\end{eqnarray}

Next we use GRV94 parton distributions \cite{GRVARXIV55A} for determining the cross-section numerically. We use Mathematica 4.1 version to carry out numerical integration with appropriate limits. To parametrize LO parton distributions as per radiative (dynamical) predictions, one has to define 

\begin{equation}
\label{eqn:ch6eq60AAAAAA}
{ s \equiv \log\left[\displaystyle{\frac{\log\displaystyle{\left[\frac{Q^{2}}{(0.232 \,\,GeV)^{2}}\right]}}{\log\displaystyle{\left[\frac{\mu_{LO}^{2}}{(0.232 \,\,GeV)^{2}}\right]}}}\right]}.
\end{equation}

to be evaluated for $\mu_{LO}^2=0.23 $ GeV$^2$. The validity of all the following parametrizations are in the range $0.4 \le Q^{2} \le 10^{6}$ GeV$^2$ (i.e. $0.3 \le s \le 2.4$) and $10^{-5} \le x< 1 $.

The non-singlet structure functions can be parametrized as
\begin{equation}
\label{eqn:ch6eq60A}
{xv(x,Q^2)=Nx^{a}(1+Ax^{b}+Bx+Cx^{\frac{3}{2}})(1-x)^{D}}.
\end{equation}

For $v=u_v$,

\begin{eqnarray}
\label{eqn:ch6eq61A}
a=0.590-0.024s,\,\, b=0.131+0.063s,\nonumber \\
N=2.284+0.802s+0.055s^{2},\nonumber \\
A=-0.449-0.138s-0.076s^{2},\nonumber \\
B=0.213+2.669s-0.728s^{2},\nonumber \\
C=8.854-9.135s+1.979s^{2},\nonumber \\
D=2.997+0.753s-0.076s^{2}.
\end{eqnarray}

for $v=d_v$

\begin{eqnarray}
\label{eqn:ch6eq62A}
a=0.376,\,\, b=0.486+0.062s,\nonumber \\
N=0.371+0.083s+0.039s^{2},\nonumber \\
A=-0.509+3.310s-1.248s^{2},\nonumber \\
B=12.41-10.52s+2.267s^{2},\nonumber \\
C=6.373-6.208s+1.418s^{2},\nonumber \\
D=3.691+0.799s-0.071s^{2}
\end{eqnarray}

The gluon and sea $\overline{u}+\overline{d}$ distributions are parametrised as
\begin{eqnarray}
\label{eqn:ch6eq63A}
xw(x,Q^2)&=&\left[x^{a}\left(A+Bx+Cx^{2}\right)+\left(\log \frac{1}{x}\right)^b+s^{\alpha} \right. \nonumber \\
& & \left. \times \exp \left(-E+\sqrt{E^{\prime}s^{\beta}\log \frac{1}{x}}\right)\right](1-x)^D.
\end{eqnarray}
where for $w=\overline{u}+\overline{d}$

\begin{eqnarray}
\label{eqn:ch6eq64A}
\alpha=1.451,\,\, \beta=0.271,\nonumber \\
a=0.410-0.232s,\,\, b=0.534-0.457s,\nonumber \\
A=0.890-0.140s,\,\, B=-0.981,\nonumber \\
C=0.320+0.683s,\nonumber \\
D=4.752+1.164s+0.286s^{2},\nonumber \\
E=4.119+1.713s,\,\,E^{\prime}=0.682+2.978s.
\end{eqnarray}

Parametrization of the strange sea distribution is as follows:

\begin{eqnarray}
\label{eqn:ch6eq65A}
xw^{\prime}(x,Q^2)&=&\frac{s^{\alpha}}{\left(\log \frac{1}{x}\right)^{a}}\left(1+A \sqrt{x}+Bx \right)(1-x)^D \nonumber\\
& &  \times \exp\left(-E+\sqrt{E^{\prime}s^{\beta}\log \frac{1}{x}}\right)
\end{eqnarray}

i.e for $w^{\prime}=s=\overline{s}$

\begin{eqnarray}
\label{eqn:ch6eq66A}
\alpha=0.914,\,\, \beta=0.577,\nonumber \\
a=1.798-0.596s,\nonumber \\
A=-5.548+3.669\sqrt{s}-0.616s,\nonumber \\
B=18.92-16.73\sqrt{s}+5.168s,\nonumber \\
D=6.379-0.350s+0.142s^{2},\nonumber \\
E=3.981+1.638s,\,\,E^{\prime}=6.402.
\end{eqnarray}

Putting these distributions in eq. (\ref{eqn:ch6eq27BB}) and eq. (\ref{eqn:ch6eq35ROMAN}), we obtain the values of cross-sections for $\nu N$ (as recorded in Table \ref{table:ch6tab3}) and $\overline{\nu} N$ (as recorded in Table \ref{table:ch6tab4}) interactions both for charged current  and neutral current  respectively. As in our analytical case, the following power law forms indicate the cross-section versus energy relation: 

\begin{equation}
\label{eqn:ch6eq44numerical}
{\sigma_{CC}^{\nu N}=\left\{{\begin{array}{r@{\quad:\quad}l}1.05 \times 10^{-36}\,\,cm^{2}\displaystyle{\left(\frac{E_{\nu}}{1\,\,GeV}\right)^{0.433}} &  10^{5}\le E_{\nu} \le 10^{8} \,GeV \\  3 \times 10^{-36}\,\,cm^{2}\displaystyle{\left(\frac{E_{\nu}}{1\,GeV}\right)^{0.383}} &  10^{8}\le E_{\nu} \le 10^{12} \,GeV \end{array}} \right.}
\end{equation}

\begin{equation}
\label{eqn:ch6eq45numerical}
{\sigma_{NC}^{\nu N}=\left\{{\begin{array}{r@{\quad:\quad}l}2.34 \times 10^{-37}\,\,cm^{2}\displaystyle{\left(\frac{E_{\nu}}{1\,\,GeV}\right)^{0.426}} &  10^{5}\le E_{\nu} \le 10^{8} \,GeV \\  5.68 \times 10^{-37}\,\,cm^{2}\displaystyle{\left(\frac{E_{\nu}}{1\,GeV}\right)^{0.387}} &  10^{8}\le E_{\nu} \le 10^{12} \,GeV \end{array}} \right.}
\end{equation}

\begin{equation}
\label{eqn:ch6eq46numerical}
{\sigma_{CC}^{\overline{\nu}N}=\left\{{\begin{array}{r@{\quad:\quad}l}6.05 \times 10^{-37}\,\,cm^{2}\displaystyle{\left(\frac{E_{\overline{\nu}}}{1\,\,GeV}\right)^{0.466}} &  10^{5}\le E_{\overline{\nu}} \le 10^{8} \,GeV \\  3 \times 10^{-36}\,\,cm^{2}\displaystyle{\left(\frac{E_{\overline{\nu}}}{1\,GeV}\right)^{0.383}} &  10^{8}\le E_{\overline{\nu}} \le 10^{12} \,GeV \end{array}} \right.}
\end{equation}

\begin{equation}
\label{eqn:ch6eq47numerical}
{\sigma_{NC}^{\overline{\nu}N}=\left\{{\begin{array}{r@{\quad:\quad}l}7.90 \times 10^{-38}\,\,cm^{2}\displaystyle{\left(\frac{E_{\overline{\nu}}}{1\,\,GeV}\right)^{0.487}} &  10^{5}\le E_{\overline{\nu}} \le 10^{8} \,GeV \\  5.68 \times 10^{-37}\,\,cm^{2}\displaystyle{\left(\frac{E_{\overline{\nu}}}{1\,GeV}\right)^{0.387}} &  10^{8}\le E_{\overline{\nu}} \le 10^{12} \,GeV \end{array}} \right.}
\end{equation}

This is to be compared with NLO expressions for cross-sections of Ref. \cite{GKR}, calculated on the basis of GRV 98 parton distribution \cite{GRVARXIV55}

\begin{equation}
\label{eqn:ch6eq44}
{\sigma_{CC}^{\nu N}=\left\{{\begin{array}{r@{\quad:\quad}l}1.10 \times 10^{-36}\,\,cm^{2}\displaystyle{\left(\frac{E_{\nu}}{1\,\,GeV}\right)^{0.454}} &  10^{5}\le E_{\nu} \le 10^{8} \,GeV \\  5.20 \times 10^{-36}\,\,cm^{2}\displaystyle{\left(\frac{E_{\nu}}{1\,GeV}\right)^{0.372}} &  10^{8}\le E_{\nu} \le 10^{12} \,GeV \end{array}} \right.}
\end{equation}

\begin{equation}
\label{eqn:ch6eq45}
{\sigma_{NC}^{\nu N}=\left\{{\begin{array}{r@{\quad:\quad}l}3.55 \times 10^{-37}\,\,cm^{2}\displaystyle{\left(\frac{E_{\nu}}{1\,\,GeV}\right)^{0.467}} &  10^{5}\le E_{\nu} \le 10^{8} \,GeV \\  3.14 \times 10^{-36}\,\,cm^{2}\displaystyle{\left(\frac{E_{\nu}}{1\,GeV}\right)^{0.349}} &  10^{8}\le E_{\nu} \le 10^{12} \,GeV \end{array}} \right.}
\end{equation}

\begin{equation}
\label{eqn:ch6eq46}
{\sigma_{CC}^{\overline{\nu}N}=\left\{{\begin{array}{r@{\quad:\quad}l}6.55 \times 10^{-37}\,\,cm^{2}\displaystyle{\left(\frac{E_{\overline{\nu}}}{1\,\,GeV}\right)^{0.484}} &  10^{5}\le E_{\overline{\nu}} \le 10^{8} \,GeV \\  5.20 \times 10^{-36}\,\,cm^{2}\displaystyle{\left(\frac{E_{\overline{\nu}}}{1\,GeV}\right)^{0.372}} &  10^{8}\le E_{\overline{\nu}} \le 10^{12} \,GeV \end{array}} \right.}
\end{equation}

\begin{equation}
\label{eqn:ch6eq47}
{\sigma_{NC}^{\overline{\nu}N}=\left\{{\begin{array}{r@{\quad:\quad}l}3.04 \times 10^{-37}\,\,cm^{2}\displaystyle{\left(\frac{E_{\overline{\nu}}}{1\,\,GeV}\right)^{0.474}} &  10^{5}\le E_{\overline{\nu}} \le 10^{8} \,GeV \\  3.14 \times 10^{-36}\,\,cm^{2}\displaystyle{\left(\frac{E_{\overline{\nu}}}{1\,GeV}\right)^{0.349}} &  10^{8}\le E_{\overline{\nu}} \le 10^{12} \,GeV \end{array}} \right.}
\end{equation}

Our expressions are also to be compared with the NLO expressions for cross-sections of Ref. \cite{RCMI1}, calculated on the basis of CTEQ3-DIS parton distributions \cite{LAIHL11}. These cross-sections are well represented by simple power-law form for neutrino energies in the range $ 10^{6}$ GeV $\le E_{\nu} \le 10^{12}$ GeV as given below:

\begin{equation}
\label{eqn:ch6eq48}
{\sigma_{CC}^{\nu N}=2.69 \times 10^{-36}\,\,cm^{2}\left(\frac{E_{\nu}}{1\,\,GeV}\right)^{0.402}}
\end{equation}

\begin{equation}
\label{eqn:ch6eq49}
{\sigma_{NC}^{\nu N}=1.06 \times 10^{-36}\,\,cm^{2}\left(\frac{E_{\nu}}{1\,\,GeV}\right)^{0.408}}
\end{equation}

\begin{equation}
\label{eqn:ch6eq50}
{\sigma_{CC}^{\overline{\nu}N}=2.53 \times 10^{-36}\,\,cm^{2}\left(\frac{E_{\overline{\nu}}}{1\,\,GeV}\right)^{0.404}}
\end{equation}

\begin{equation}
\label{eqn:ch6eq51}
{\sigma_{NC}^{\overline{\nu}N}=0.98 \times 10^{-36}\,\,cm^{2}\left(\frac{E_{\overline{\nu}}}{1\,\,GeV}\right)^{0.410}}
\end{equation}

Next, we compare our results with the NLO results of Ref. \cite{RCMI111}, calculated on the basis of CTEQ4-DIS parton distributions \cite{LAIHL1122}. For neutrino energies in the range $ 10^{7}$ GeV $\le E_{\nu} \le 10^{12}$ GeV , the CTEQ4-DIS cross sections are given within 10\% by 

\begin{equation}
\label{eqn:ch6eq52}
{\sigma_{CC}^{\nu N}=5.53 \times 10^{-36}\,\,cm^{2}\left(\frac{E_{\nu}}{1\,\,GeV}\right)^{0.363}}
\end{equation}

\begin{equation}
\label{eqn:ch6eq53}
{\sigma_{NC}^{\nu N}=2.31 \times 10^{-36}\,\,cm^{2}\left(\frac{E_{\nu}}{1\,\,GeV}\right)^{0.363}}
\end{equation}

\begin{equation}
\label{eqn:ch6eq54}
{\sigma_{tot}^{\nu N}=7.84 \times 10^{-36}\,\,cm^{2}\left(\frac{E_{\nu}}{1\,\,GeV}\right)^{0.363}}
\end{equation}

\begin{equation}
\label{eqn:ch6eq55}
{\sigma_{CC}^{\overline{\nu}N}=5.52 \times 10^{-36}\,\,cm^{2}\left(\frac{E_{\overline{\nu}}}{1\,\,GeV}\right)^{0.363}}
\end{equation}

\begin{equation}
\label{eqn:ch6eq56}
{\sigma_{NC}^{\overline{\nu}N}=2.29 \times 10^{-36}\,\,cm^{2}\left(\frac{E_{\overline{\nu}}}{1\,\,GeV}\right)^{0.363}}
\end{equation}

\begin{equation}
\label{eqn:ch6eq57}
{\sigma_{tot}^{\overline{\nu} N}=7.80 \times 10^{-36}\,\,cm^{2}\left(\frac{E_{\overline{\nu}}}{1\,\,GeV}\right)^{.363}}
\end{equation}

It is pertinent to note that CTEQ3-DIS parton distributions are more singular than CTEQ4-DIS parton distributions as $x$ becomes ultra small i.e. as $x\rightarrow 0$ \cite{RCMI111}. Around $x\rightarrow 0$, CTEQ3 sea-quark distributions are presented by
 
\begin{equation}
\label{eqn:ch6eq58}
{xq_{s}^{CTEQ3}(x)\propto x^{-0.332}}
\end{equation}

On the other hand, the behaviour of CTEQ4 sea-quark distributions is well-presented by

\begin{equation}
\label{eqn:ch6eq59}
{xq_{s}^{CTEQ4}(x)\propto x^{-0.227}}
\end{equation}

Next we compare our analytical and numerical results with the analytical asymptotic approximation technique of Fiore et al as in Ref. \cite{FIOREJKPPPR12} and their numerically obtained results \cite{FIOREJKPPPR12546}.

We also compare our analytical and numerical results with the numerical results of Rahul Basu et al \cite{RAHULBASU} obtained by using GRV 98 NLO parton distributions.

We next find the ratio of $\sigma_{CC}^{\nu N}$ and $\sigma_{CC}^{\overline{\nu}N}$ and the ratio comes closer to the value 1 at the higher end of the neutrino (antineutrino) energy spectrum. This is because at higher energies (consequently at lower $x$), the contribution due to $xF_{3}$ part (valence part) becomes neglible in comparison to $F_{2}$ part (sea and gluon part).

\begin{figure}[t]
\begin{center}
\includegraphics[height=14cm,width=16cm]{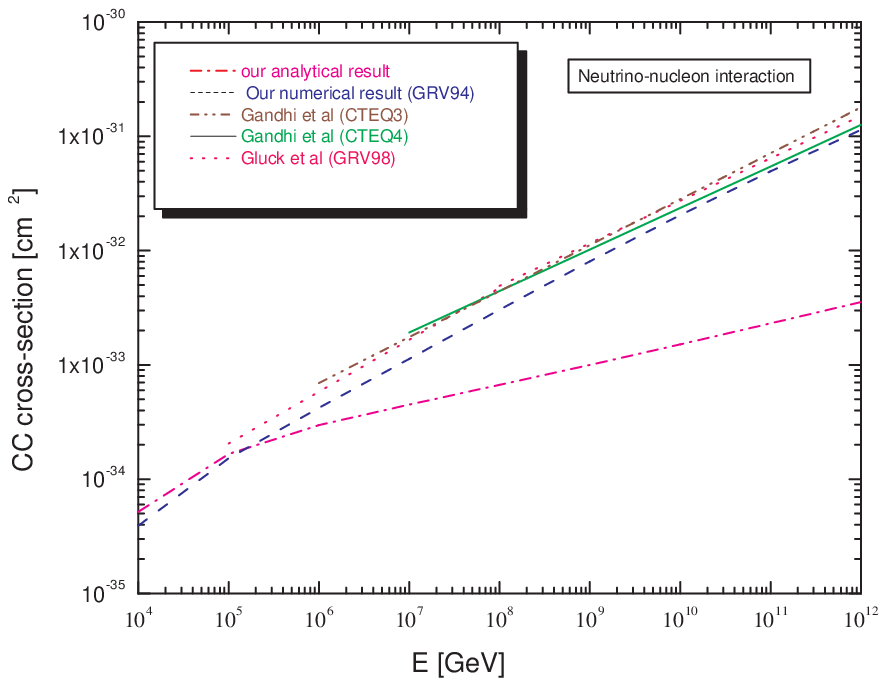}
\end{center}
\caption[UHE neutrino-nucleon charged current cross-section $\sigma_{CC}^{\nu N}\,(cm^{2})$ obtained from our analytic LO expression. This is compared with our numerical LO expression. In the same figure, NLO results of Ref. \protect\cite{GKR} and that of Ref. \protect\cite{RCMI1} as well as Ref. \protect\cite{RCMI111} are also plotted.]{UHE neutrino-nucleon charged current cross-section $\sigma_{CC}^{\nu N}\,(cm^{2})$ obtained from our analytic LO expression eq. (\ref{eqn:ch6eq44analytical}) (Magenta coloured dash dot line). This is compared with our numerical LO expression eq. (\ref{eqn:ch6eq44numerical}) (Blue coloured dash line). In the same figure, we also plot NLO results of Ref. \protect\cite{GKR} eq. (\ref{eqn:ch6eq44}) (Pink coloured dot line) and that of Ref. \protect\cite{RCMI111} (Green coloured solid line) eq. (\ref{eqn:ch6eq48}) as well as Ref. \cite{RCMI1} eq. (\ref{eqn:ch6eq52}) (Wine coloured dash dot dot line)}. 
\label{fig:ch6fig1}
\end{figure}

\begin{figure}[t]
\begin{center}
\includegraphics[height=14cm,width=16cm]{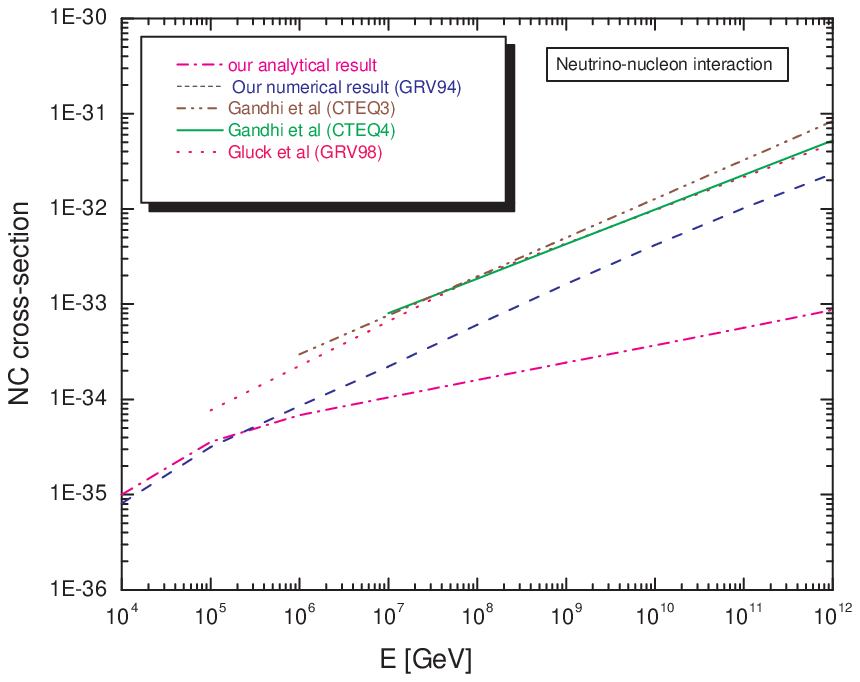}
\end{center}
\caption[UHE neutrino-nucleon neutral current cross-section $\sigma_{NC}^{\nu N}\,(cm^{2})$ obtained from our analytic LO expression. This is compared with our numerical LO expression. In the same figure, NLO results of Ref. \protect\cite{GKR} and that of Ref. \protect\cite{RCMI1} as well as Ref. \protect\cite{RCMI111} are also plotted.]{UHE neutrino-nucleon neutral current cross-section $\sigma_{NC}^{\nu N}\,(cm^{2})$ obtained from our analytic LO expression eq. (\ref{eqn:ch6eq45analytical}) (Magenta coloured dash dot line). This is compared with our numerical LO expression eq. (\ref{eqn:ch6eq45numerical}) (Blue coloured dash line). In the same figure, we also plot NLO results of Ref. \protect\cite{GKR} eq. (\ref{eqn:ch6eq45}) (Pink coloured dot line) and that of Ref. \cite{RCMI111} (Green coloured solid line) eq. (\ref{eqn:ch6eq49}) as well as Ref. \protect\cite{RCMI1} eq. (\ref{eqn:ch6eq53}) (Wine coloured dash dot dot line)}.
\label{fig:ch6fig2}
\end{figure}

\begin{figure}[t]
\begin{center}
\includegraphics[height=14cm,width=16cm]{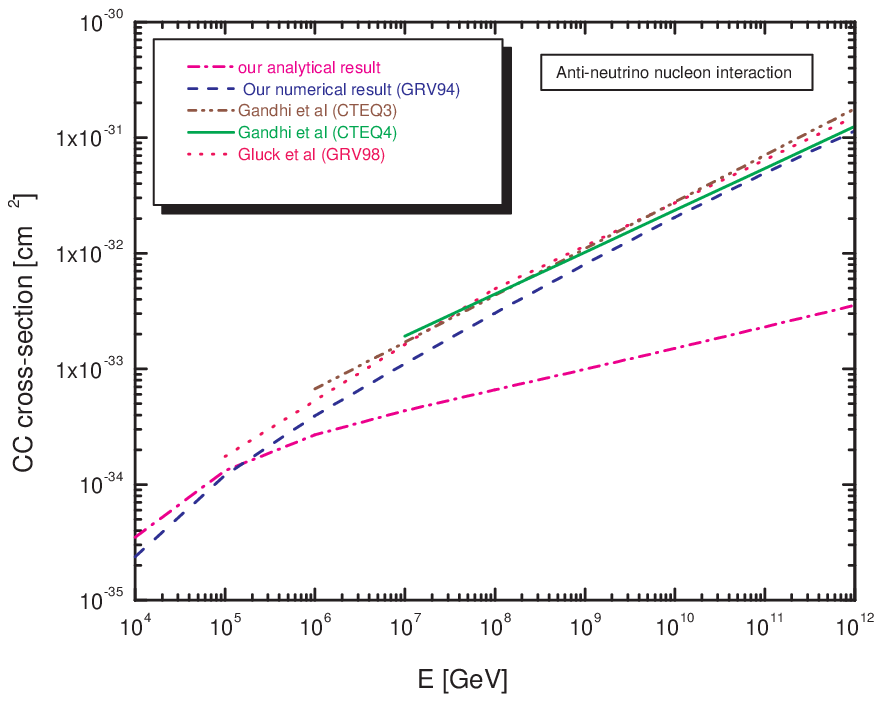}
\end{center}
\caption[UHE anti-neutrino nucleon charged current cross-section $\sigma_{CC}^{\overline{\nu} N}\,(cm^{2})$ obtained from our analytic LO expression. This is compared with our numerical LO expression. In the same figure, NLO results of Ref. \protect\cite{GKR} and that of Ref. \protect\cite{RCMI1} as well as Ref. \protect\cite{RCMI111} are also plotted.]
{UHE anti-neutrino nucleon charged current cross-section $\sigma_{CC}^{\overline{\nu} N}\,(cm^{2})$ obtained from our analytic LO expression eq. (\ref{eqn:ch6eq46analytical}) (Magenta coloured dash dot line). This is compared with our numerical LO expression eq. (\ref{eqn:ch6eq46numerical}) (Blue coloured dash line). In the same figure, we also plot NLO results of Ref. \protect\cite{GKR} eq. (\ref{eqn:ch6eq46}) (Pink coloured dot line) and that of Ref. \cite{RCMI111} (Green coloured solid line) eq. (\ref{eqn:ch6eq50}) as well as Ref. \protect\cite{RCMI1} eq. (\ref{eqn:ch6eq55}) (Wine coloured dash dot dot line).} 
\label{fig:ch6fig3}
\end{figure}

\begin{figure}[t]
\begin{center}
\includegraphics[height=14cm,width=16cm]{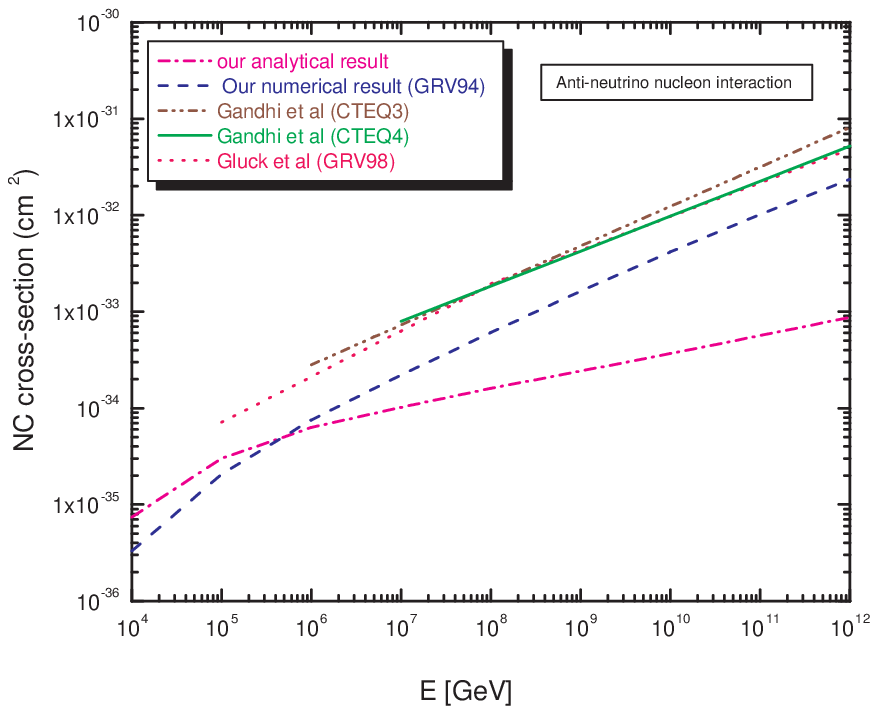}
\end{center}
\caption[UHE anti-neutrino nucleon neutral current cross-section $\sigma_{NC}^{\overline{\nu} N}\,(cm^{2})$  versus $E_{\nu}$ obtained from our analytic LO expression. This is compared with our numerical LO expression. In the same figure, NLO results of Ref. \protect\cite{GKR} and that of Ref. \protect\cite{RCMI1} as well as Ref. \protect\cite{RCMI111} are also plotted.]{UHE anti-neutrino nucleon neutral current cross-section $\sigma_{NC}^{\overline{\nu} N}\,(cm^{2})$ obtained from our analytic LO expression eq. (\ref{eqn:ch6eq47analytical}) (Magenta coloured dash dot line). This is compared with our numerical LO expression eq. (\ref{eqn:ch6eq47numerical}) (Blue coloured dash line). In the same figure, we also plot NLO results of Ref. \protect\cite{GKR} eq. (\ref{eqn:ch6eq47}) (Pink coloured dot line) and that of Ref. \cite{RCMI111} (Green coloured solid line) eq. (\ref{eqn:ch6eq51}) as well as Ref. \protect\cite{RCMI1} eq. (\ref{eqn:ch6eq56}) (Wine coloured dash dot dot line).} 
\label{fig:ch6fig4}
\end{figure}

\begin{figure}[t]
\begin{center}
\includegraphics[height=14cm,width=16cm]{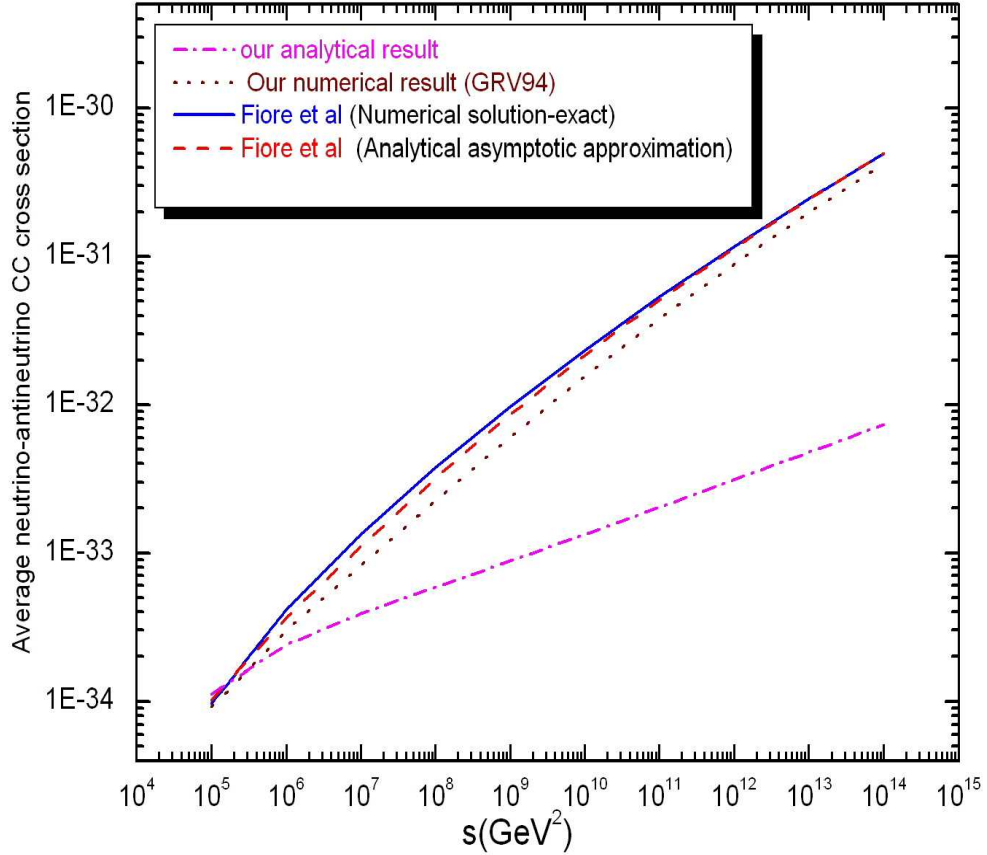}
\end{center}
\caption[Comparison is done between average neutrino(antineutrino)-nucleon charged current cross section in cm$^{2}$ according to the numerical results obtained in Ref. \protect\cite{FIOREJKPPPR12546}, the asymptotic approximation method of Ref. \protect\cite{FIOREJKPPPR12}, our analytical evaluation obtained using eq. (\ref{eqn:ch6eq44analytical}) as well as eq. (\ref{eqn:ch6eq46analytical}) and our numerical evaluation obtained using eq. (\ref{eqn:ch6eq44numerical}) as well as eq. (\ref{eqn:ch6eq46numerical})]{We compare between average neutrino(antineutrino)-nucleon charged current cross section in cm$^{2}$ according to the numerical results obtained in Ref. \protect\cite{FIOREJKPPPR12546} (Blue coloured solid line), the asymptotic approximation method of Ref. \protect\cite{FIOREJKPPPR12} (Red coloured dash line), our analytical evaluation (Magenta coloured dash dot line) obtained using eq. (\ref{eqn:ch6eq44analytical}) as well as eq. (\ref{eqn:ch6eq46analytical}) and our numerical evaluation obtained (Wine coloured dot line) using eq. (\ref{eqn:ch6eq44numerical}) as well as eq. (\ref{eqn:ch6eq46numerical})}
\label{fig:ch6fig5}
\end{figure}

\begin{figure}[t]
\begin{center}
\includegraphics[height=14cm,width=16cm]{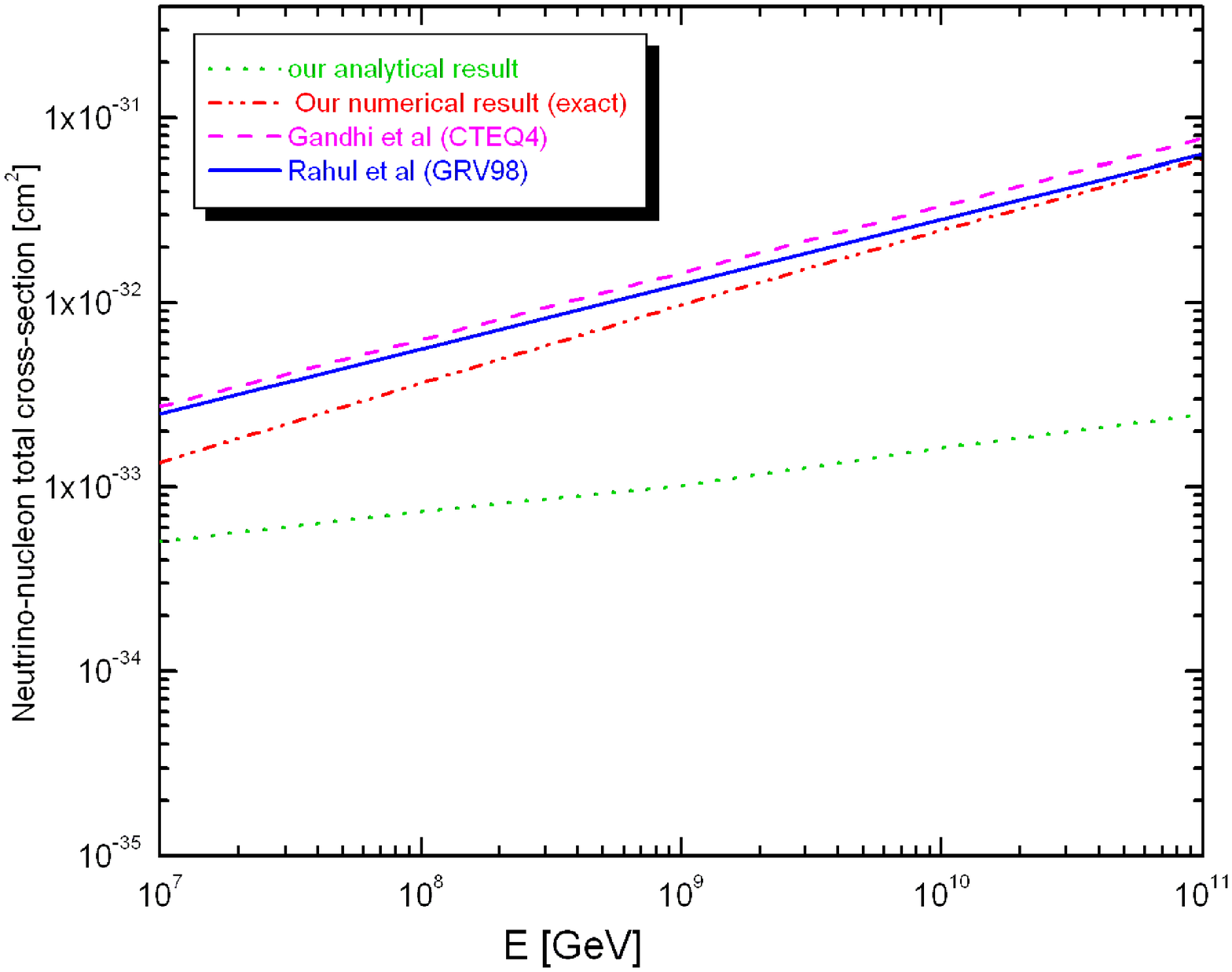}
\end{center}
\caption[ Comparison is done between the neutrino-nucleon total cross sections in cm$^{2}$ according to the numerical results obtained in Ref. \protect\cite{RAHULBASU}, the numerical method of Ref. \protect\cite{RCMI111}, our analytical evaluation and our numerical evaluation.]{We plot  the neutrino-nucleon total cross sections in cm$^{2}$ versus neutrino energy to compare the numerical results obtained in Ref. \protect\cite{RAHULBASU} (Blue coloured solid line), the numerical method of Ref. \protect\cite{RCMI111} (Magenta coloured dash line), our analytical evaluation obtained using eq. (\ref{eqn:ch6eq44analytical}) (Olive coloured dot line) as well as eq. (\ref{eqn:ch6eq45analytical}) and our numerical evaluation obtained using eq. (\ref{eqn:ch6eq44numerical}) (Red coloured dash dot dot line) as well as eq. (\ref{eqn:ch6eq45numerical})}
\label{fig:ch6fig6}
\end{figure}

 \begin{figure}[t]
\begin{center}
\includegraphics[height=14cm,width=16cm]{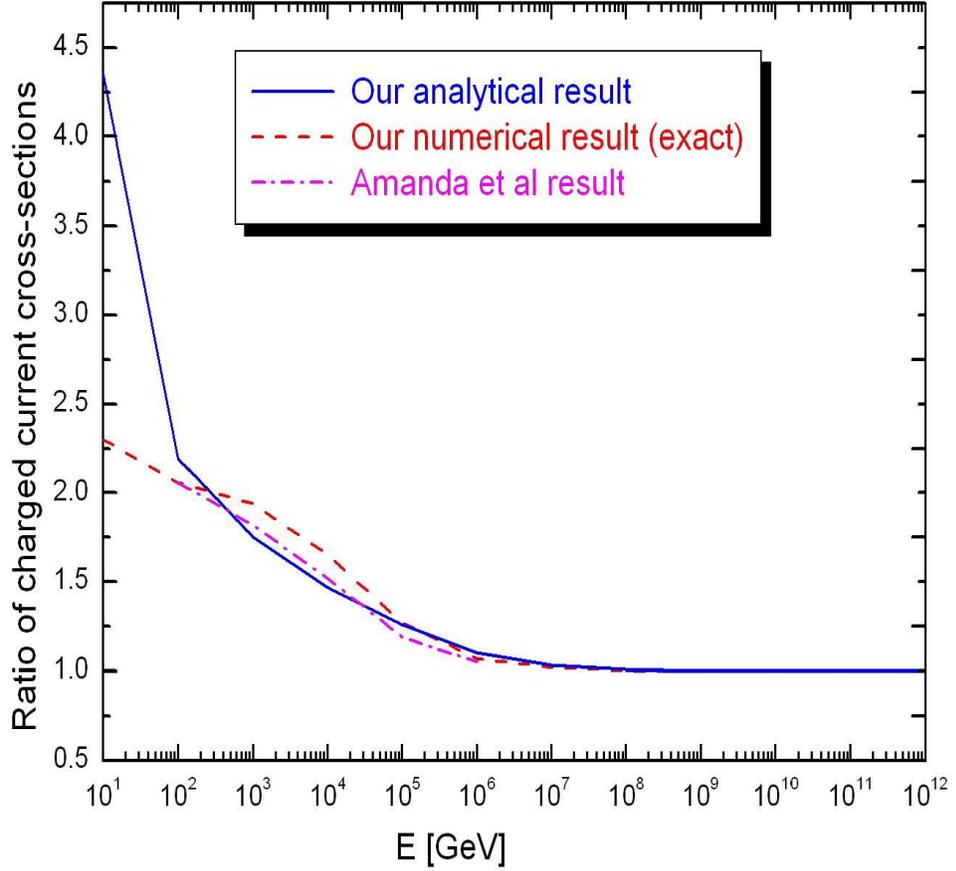}
\end{center}
\caption[The ratio of neutrino-nucleon charged current cross-section and antineutrino-nucleon charged current cross-section $\left(\displaystyle{\frac{\sigma_{CC}^{\nu N}}{\sigma_{CC}^{\overline{\nu}N}}}\right)$ of our analytical and numerical method versus neutrino (antineutrino) energy $\left(E_{\nu(\overline{\nu})}\right)$.]{We plot the ratio of neutrino-nucleon charged current cross-section and antineutrino-nucleon charged current cross-section $\left(\frac{\sigma_{CC}^{\nu N}}{\sigma_{CC}^{\overline{\nu}N}}\right)$ of our analytical (Table 6.1 and 6.2) (Blue coloured solid line) and numerical method (Table 6.3 and 6.4) (Red coloured dash line). The ratio obtained from Ref \protect\cite{ACSSS} (Magenta coloured dash dot line) versus neutrino (antineutrino) energy $\left(E_{\nu(\overline{\nu})}\right)$ is also plotted for comparison.}
\label{fig:ch6fig7}
\end{figure}

\begin{figure}[t]
\begin{center}
\includegraphics[height=14cm,width=16cm]{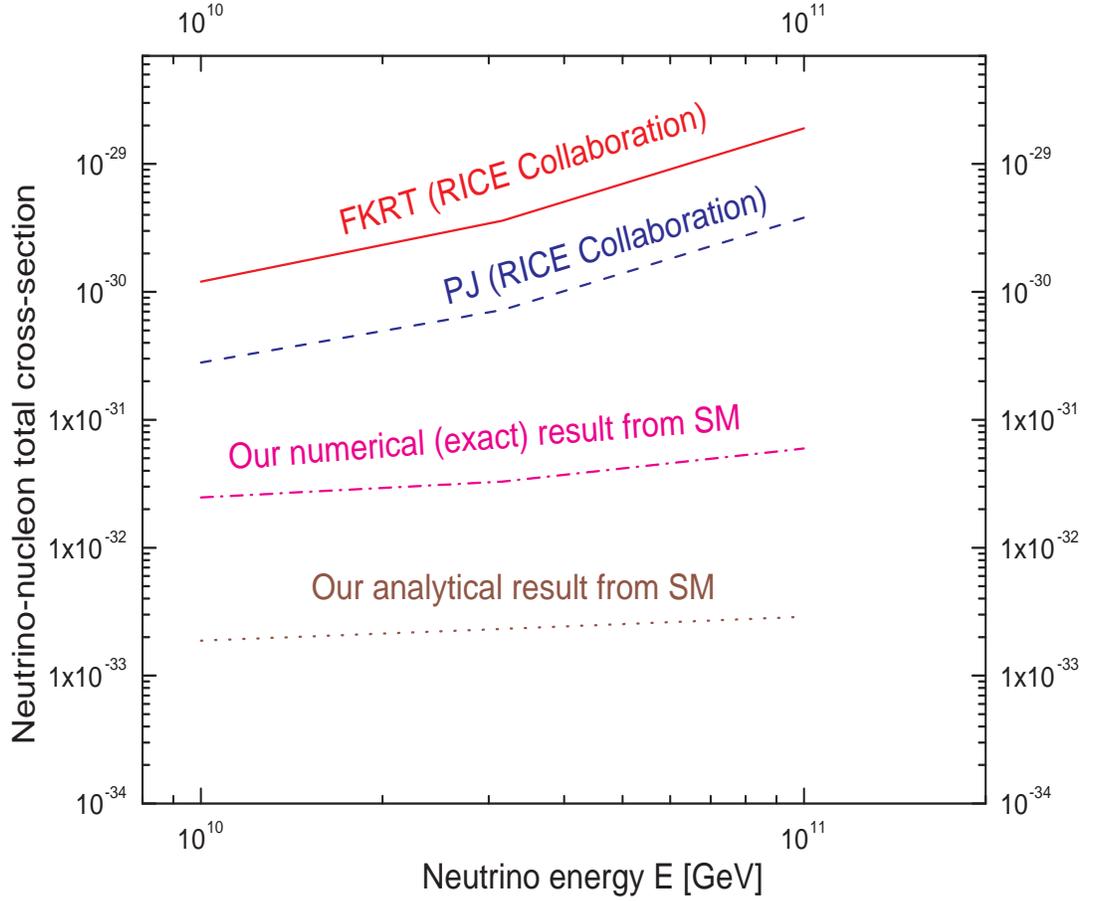}
\end{center}
\caption[Model independent upper bounds on the neutrino-nucleon inelastic total cross-sections derived from the RICE \protect\cite{Krai,IKRAVCHENCO} collaboration search results by exploiting the cosmogenic neutrino flux estimates of Ref. \protect\cite{ZFODORSDKARHTU} and Ref. \protect\cite{RJPROPAJOHN} are plotted against neutrino energy. Our analytical as well as numerically (exact) determined results from Standard Model for the Ultra High Energy neutrino-nucleon total cross-section in leading order are compared with the above-mentioned upper bounds at three different neutrino energies $E_\nu = 10^{10}, 10^{10.5}, 10^{11}$ GeV.]{Model independent upper bounds on the neutrino-nucleon inelastic total cross-sections derived from the RICE \protect\cite{Krai,IKRAVCHENCO} collaboration search results by exploiting the cosmogenic neutrino flux estimates of Ref. \protect\cite{ZFODORSDKARHTU} and Ref. \protect\cite{RJPROPAJOHN} are plotted against neutrino energy. Our analytical as well as numerically (exact) determined results from Standard Model for the Ultra High Energy neutrino-nucleon total cross-section in leading order are compared with the above-mentioned upper bounds at three different neutrino energies $E_\nu = 10^{10}, 10^{10.5}, 10^{11}$ GeV.}
\label{fig:Trombay}
\end{figure}

\begin{table}
\begin{center}
\vspace{-0.3in}
\begin{tabular}{|c|c|c|c|}\hline
\multicolumn{4}{|c|}{\rule[-.3cm]{0mm}{1.4cm}\bfseries  \Large{\bf{Energy cum $\nu N$ interaction cross-section}}} \\ \hline {\rule[-.3cm]{0mm}{1.2cm}}
\large{$E_{\nu}$\,\,(GeV)}  & \large{$\sigma_{CC}^{\nu N}$\,(cm$^{2}$)} & \large{$\sigma_{NC}^{\nu N}$\,(cm$^{2}$)} &  \large{$\sigma_{tot}^{\nu N}$\,(cm$^{2}$)} \\ \hline {\rule[-.3cm]{0mm}{1cm}}
$10$ 	        & $4.36\times10^{-38}$	     & $7.24\times10^{-39}$  & $5.08\times10^{-38}$\\ \hline {\rule[-.3cm]{0mm}{1cm}}
$10^{2}$	& $6.22\times10^{-37}$	     & $1.08\times10^{-37}$  & $7.30\times10^{-37}$\\ \hline {\rule[-.3cm]{0mm}{.8cm}}
$10^{3}$	& $6.82\times10^{-36}$	     & $1.22\times10^{-36}$  & $8.04\times10^{-36}$\\ \hline {\rule[-.3cm]{0mm}{.8cm}}
$10^{4}$	& $5.17\times10^{-35}$	     & $1.00\times10^{-35}$  & $6.17\times10^{-35}$\\ \hline {\rule[-.3cm]{0mm}{.8cm}}
$10^{5}$	& $1.66\times10^{-34}$	     & $3.59\times10^{-35}$  & $2.02\times10^{-34}$\\ \hline {\rule[-.3cm]{0mm}{.8cm}}
$10^{6}$	& $2.97\times10^{-34}$	     & $6.82\times10^{-35}$  & $3.65\times10^{-34}$\\ \hline {\rule[-.3cm]{0mm}{.8cm}}
$10^{7}$	& $4.50\times10^{-34}$	     & $1.05\times10^{-34}$  & $5.55\times10^{-34}$\\ \hline {\rule[-.3cm]{0mm}{.8cm}}
$10^{8}$	& $6.68\times10^{-34}$	     & $1.60\times10^{-34}$  & $8.28\times10^{-34}$\\ \hline {\rule[-.3cm]{0mm}{.8cm}}
$10^{9}$	& $1\times10^{-33}$	     & $2.43\times10^{-34}$  & $1.24\times10^{-33}$\\ \hline {\rule[-.3cm]{0mm}{.8cm}}
$10^{10}$	& $1.51\times10^{-33}$	     & $3.70\times10^{-34}$  & $1.88\times10^{-33}$\\ \hline {\rule[-.3cm]{0mm}{.8cm}}
$10^{11}$	& $2.31\times10^{-33}$	     & $5.66\times10^{-34}$  & $2.88\times10^{-33}$\\ \hline {\rule[-.3cm]{0mm}{.8cm}}
$10^{12}$	& $3.54\times10^{-33}$	     & $8.71\times10^{-34}$  & $4.41\times10^{-33}$\\ \hline 

\end{tabular}
\vspace{0.6in}
\caption{Analytical estimation of neutrino-induced charged current, neutral current and total cross-sections in cm$^{2}$ on the basis of eq. (\ref{eqn:ch6eq44analytical}) and eq. (\ref{eqn:ch6eq45analytical}), considering the simple ansatz for the MRST f4 2004 input parton distributions \protect\cite{MRST2222,MRSTDURHAM}}
\label{table:ch6tab1}
\end{center}
\end{table}

\begin{table}
\begin{center}
\vspace{-0.3in}
\begin{tabular}{|c|c|c|c|}\hline
\multicolumn{4}{|c|}{\rule[-.3cm]{0mm}{1.4cm}\bfseries  \Large{\bf{Energy cum $\overline{\nu} N$ interaction cross-section}}} \\ \hline {\rule[-.3cm]{0mm}{1.2cm}}
\large{$E_{\overline{\nu}}$\,\,(GeV)}  & \large{$\sigma_{CC}^{\overline{\nu} N}$\,(cm$^{2}$)} & \large{$\sigma_{NC}^{\overline{\nu} N}$\,(cm$^{2}$)} &  \large{$\sigma_{tot}^{\overline{\nu} N}$\,(cm$^{2}$)} \\ \hline {\rule[-.3cm]{0mm}{1cm}}
$10$ 	        & $1.00\times10^{-38}$	     & $2.46\times10^{-39}$  & $1.25\times10^{-38}$\\ \hline {\rule[-.3cm]{0mm}{1cm}}
$10^{2}$	& $2.84\times10^{-37}$	     & $6.01\times10^{-38}$  & $3.44\times10^{-37}$\\ \hline {\rule[-.3cm]{0mm}{.8cm}}
$10^{3}$	& $3.90\times10^{-36}$	     & $8\times10^{-37}$     & $4.70\times10^{-36}$\\ \hline {\rule[-.3cm]{0mm}{.8cm}}
$10^{4}$	& $3.51\times10^{-35}$	     & $7.40\times10^{-36}$  & $4.25\times10^{-35}$\\ \hline {\rule[-.3cm]{0mm}{.8cm}}
$10^{5}$	& $1.32\times10^{-34}$	     & $3.01\times10^{-35}$  & $1.62\times10^{-34}$\\ \hline {\rule[-.3cm]{0mm}{.8cm}}
$10^{6}$	& $2.69\times10^{-34}$	     & $6.30\times10^{-35}$  & $3.32\times10^{-34}$\\ \hline {\rule[-.3cm]{0mm}{.8cm}}
$10^{7}$	& $4.35\times10^{-34}$	     & $1.02\times10^{-34}$  & $5.37\times10^{-34}$\\ \hline {\rule[-.3cm]{0mm}{.8cm}}
$10^{8}$	& $6.61\times10^{-34}$	     & $1.60\times10^{-34}$  & $8.21\times10^{-34}$\\ \hline {\rule[-.3cm]{0mm}{.8cm}}
$10^{9}$	& $9.96\times10^{-34}$	     & $2.43\times10^{-34}$  & $1.24\times10^{-33}$\\ \hline {\rule[-.3cm]{0mm}{.8cm}}
$10^{10}$	& $1.51\times10^{-33}$	     & $3.70\times10^{-34}$  & $1.88\times10^{-33}$\\ \hline {\rule[-.3cm]{0mm}{.8cm}}
$10^{11}$	& $2.31\times10^{-33}$	     & $5.66\times10^{-34}$  & $2.88\times10^{-33}$\\ \hline {\rule[-.3cm]{0mm}{.8cm}}
$10^{12}$	& $3.54\times10^{-33}$	     & $8.71\times10^{-34}$  & $4.41\times10^{-33}$\\ \hline 

\end{tabular}
\vspace{0.6in}
\caption{Analytical estimation of antineutrino-induced charged current, neutral current and total cross-sections in cm$^{2}$ on the basis of eq. (\ref{eqn:ch6eq46analytical}) and eq. (\ref{eqn:ch6eq47analytical}), considering the simple ansatz for the MRST f4 2004 input parton distributions \protect\cite{MRST2222,MRSTDURHAM}}
\label{table:ch6tab2}
\end{center}
\end{table}

\begin{table}
\begin{center}
\vspace{-0.3in}
\begin{tabular}{|c|c|c|c|}\hline
\multicolumn{4}{|c|}{\rule[-.3cm]{0mm}{1.4cm}\bfseries  \Large{\bf{Energy cum $\nu N$ interaction cross-section}}} \\ \hline {\rule[-.3cm]{0mm}{1.2cm}}
\large{$E_{\nu}$\,\,(GeV)}  & \large{$\sigma_{CC}^{\nu N}$\,(cm$^{2}$)} & \large{$\sigma_{NC}^{\nu N}$\,(cm$^{2}$)} &  \large{$\sigma_{tot}^{\nu N}$\,(cm$^{2}$)} \\ \hline {\rule[-.3cm]{0mm}{1cm}}
$10$ 	        & $6.83\times10^{-38}$	     & $1.40\times10^{-38}$  & $8.23\times10^{-38}$\\ \hline {\rule[-.3cm]{0mm}{1cm}}
$10^{2}$	& $6.52\times10^{-37}$	     & $1.32\times10^{-37}$  & $7.84\times10^{-37}$\\ \hline {\rule[-.3cm]{0mm}{.8cm}}
$10^{3}$	& $5.74\times10^{-36}$	     & $1.16\times10^{-36}$  & $6.90\times10^{-36}$\\ \hline {\rule[-.3cm]{0mm}{.8cm}}
$10^{4}$	& $3.91\times10^{-35}$	     & $8.10\times10^{-36}$  & $4.72\times10^{-35}$\\ \hline {\rule[-.3cm]{0mm}{.8cm}}
$10^{5}$	& $1.52\times10^{-34}$	     & $3.16\times10^{-35}$  & $1.84\times10^{-34}$\\ \hline {\rule[-.3cm]{0mm}{.8cm}}
$10^{6}$	& $4.21\times10^{-34}$	     & $8.53\times10^{-35}$  & $5.06\times10^{-34}$\\ \hline {\rule[-.3cm]{0mm}{.8cm}}
$10^{7}$	& $1.13\times10^{-33}$	     & $2.21\times10^{-34}$  & $1.35\times10^{-33}$\\ \hline {\rule[-.3cm]{0mm}{.8cm}}
$10^{8}$	& $3.04\times10^{-33}$	     & $6.06\times10^{-34}$  & $3.65\times10^{-33}$\\ \hline {\rule[-.3cm]{0mm}{.8cm}}
$10^{9}$	& $8.07\times10^{-33}$	     & $1.63\times10^{-33}$  & $9.70\times10^{-33}$\\ \hline {\rule[-.3cm]{0mm}{.8cm}}
$10^{10}$	& $2.05\times10^{-32}$	     & $4.17\times10^{-33}$  & $2.47\times10^{-32}$\\ \hline {\rule[-.3cm]{0mm}{.8cm}}
$10^{11}$	& $4.95\times10^{-32}$	     & $1.02\times10^{-32}$  & $5.97\times10^{-32}$\\ \hline {\rule[-.3cm]{0mm}{.8cm}}
$10^{12}$	& $1.14\times10^{-31}$	     & $2.35\times10^{-32}$  & $1.38\times10^{-31}$\\ \hline 

\end{tabular}
\vspace{0.6in}
\caption{Numerical estimation of neutrino-induced charged current, neutral current and total cross-sections in cm$^{2}$ on the basis of eq. (\ref{eqn:ch6eq44numerical}) and eq. (\ref{eqn:ch6eq45numerical}), considering the GRV 94 parton distributions (which are dependent on both $x$ and $Q^{2}$) \protect\cite{GRVARXIV55A}}
\label{table:ch6tab3}
\end{center}
\end{table}

\begin{table}
\begin{center}
\vspace{-0.3in}
\begin{tabular}{|c|c|c|c|}\hline
\multicolumn{4}{|c|}{\rule[-.3cm]{0mm}{1.4cm}\bfseries  \Large{\bf{Energy cum $\overline{\nu} N$ interaction cross-section}}} \\ \hline {\rule[-.3cm]{0mm}{1.2cm}}
\large{$E_{\overline{\nu}}$\,\,(GeV)}  & \large{$\sigma_{CC}^{\overline{\nu} N}$\,(cm$^{2}$)} & \large{$\sigma_{NC}^{\overline{\nu} N}$\,(cm$^{2}$)} &  \large{$\sigma_{tot}^{\overline{\nu} N}$\,(cm$^{2}$)} \\ \hline {\rule[-.3cm]{0mm}{1cm}}
$10$ 	       & $2.97\times10^{-38}$	     & $2.98\times10^{-39}$  & $3.27\times10^{-38}$\\ \hline {\rule[-.3cm]{0mm}{1cm}}
$10^{2}$	& $3.16\times10^{-37}$	     & $3.61\times10^{-38}$  & $3.52\times10^{-37}$\\ \hline {\rule[-.3cm]{0mm}{.8cm}}
$10^{3}$	& $2.96\times10^{-36}$	     & $3.56\times10^{-37}$  & $3.32\times10^{-36}$\\ \hline {\rule[-.3cm]{0mm}{.8cm}}
$10^{4}$	& $2.37\times10^{-35}$	     & $3.30\times10^{-36}$  & $2.70\times10^{-35}$\\ \hline {\rule[-.3cm]{0mm}{.8cm}}
$10^{5}$	& $1.20\times10^{-34}$	     & $2.06\times10^{-35}$  & $1.41\times10^{-34}$\\ \hline {\rule[-.3cm]{0mm}{.8cm}}
$10^{6}$	& $3.95\times10^{-34}$	     & $7.54\times10^{-35}$  & $4.70\times10^{-34}$\\ \hline {\rule[-.3cm]{0mm}{.8cm}}
$10^{7}$	& $1.11\times10^{-33}$	     & $2.21\times10^{-34}$  & $1.33\times10^{-33}$\\ \hline {\rule[-.3cm]{0mm}{.8cm}}
$10^{8}$	& $3.04\times10^{-33}$	     & $6.06\times10^{-34}$  & $3.64\times10^{-33}$\\ \hline {\rule[-.3cm]{0mm}{.8cm}}
$10^{9}$	& $8.07\times10^{-33}$	     & $1.63\times10^{-33}$  & $9.70\times10^{-33}$\\ \hline {\rule[-.3cm]{0mm}{.8cm}}
$10^{10}$	& $2.05\times10^{-32}$	     & $4.17\times10^{-33}$  & $2.47\times10^{-32}$\\ \hline {\rule[-.3cm]{0mm}{.8cm}}
$10^{11}$	& $4.95\times10^{-32}$	     & $1.02\times10^{-32}$  & $5.97\times10^{-32}$\\ \hline {\rule[-.3cm]{0mm}{.8cm}}
$10^{12}$	& $1.14\times10^{-31}$	     & $2.35\times10^{-32}$  & $1.38\times10^{-31}$\\ \hline 

\end{tabular}
\vspace{0.6in}
\caption{Numerical estimation of antineutrino induced charged current, neutral current and total cross-sections in cm$^{2}$ on the basis of eq. (\ref{eqn:ch6eq46numerical}) and eq. (\ref{eqn:ch6eq47numerical}), considering the GRV 94 parton distributions (which are dependent on both $x$ and $Q^{2}$) \protect\cite{GRVARXIV55A}}
\label{table:ch6tab4}
\end{center}
\end{table}

\begin{table}
\begin{center}
\vspace{-0.3in}
\begin{tabular}{|c|c|c|c|c|}\hline
\multicolumn{5}{|c|}{\rule[-.3cm]{0mm}{1.4cm}\bfseries  \large{\bf{Value of $s$ cum $\frac{\nu N+ \overline{\nu} N}{2}$ interaction cross-section (cm$^{2})$}}} \\ \hline {\rule[-.3cm]{0mm}{1.2cm}}
$s$\,\,(GeV$^2$)  &  $\overline{\sigma}_{CC}^{\nu N}$, Ref. \cite{FIOREJKPPPR1} & $\overline{\sigma}_{CC}^{\nu N}$, Ref. \cite{FIOREJKPPPR12} &  $\overline{\sigma}_{CC}^{\nu N}$, analytical &   $\overline{\sigma}_{CC}^{\nu N}$, numerical   \\ \hline {\rule[-.3cm]{0mm}{1cm}}

$10^{5}$        & $9.75\times10^{-35}$	     & $1.03\times10^{-34}$  & $1.13\times10^{-34}$ & $9.25\times10^{-35}$ \\ \hline {\rule[-.3cm]{0mm}{1cm}}
$10^{6}$	& $4.13\times10^{-34}$	     &$3.65\times10^{-34}$    & $2.41\times10^{-34}$ & $2.99\times10^{-34}$ \\ \hline {\rule[-.3cm]{0mm}{.8cm}}
$10^{7}$	& $1.34\times10^{-33}$	     &$1.11\times10^{-33}$   & $3.89\times10^{-34}$ & $8.26\times10^{-34}$ \\ \hline {\rule[-.3cm]{0mm}{.8cm}}
$10^{8}$	& $3.75\times10^{-33}$	     &$3.20\times10^{-33}$    & $5.87\times10^{-34}$ & $2.25\times10^{-33}$ \\ \hline {\rule[-.3cm]{0mm}{.8cm}}
$10^{9}$	& $9.67\times10^{-33}$	     &$8.61\times10^{-33}$    & $8.82\times10^{-34}$ & $6.04\times10^{-33}$ \\ \hline {\rule[-.3cm]{0mm}{.8cm}}
$10^{10}$	& $2.33\times10^{-32}$	     &$2.16\times10^{-32}$   & $1.33\times10^{-33}$ & $1.56\times10^{-32}$ \\ \hline {\rule[-.3cm]{0mm}{.8cm}}
$10^{11}$	& $5.34\times10^{-32}$	     &$5.11\times10^{-32}$    & $2.03\times10^{-33}$ & $3.81\times10^{-32}$ \\ \hline {\rule[-.3cm]{0mm}{.8cm}}
$10^{12}$	& $1.17\times10^{-31}$	     &$1.14\times10^{-31}$   & $3.11\times10^{-33}$ & $8.89\times10^{-32}$ \\ \hline {\rule[-.3cm]{0mm}{.8cm}}
$10^{13}$	& $2.45\times10^{-31}$	     &$2.44\times10^{-31}$    & $4.78\times10^{-33}$ & $1.98\times10^{-31}$ \\ \hline {\rule[-.3cm]{0mm}{.8cm}}
$10^{14}$	& $4.97\times10^{-31}$	     &$5.02\times10^{-31}$    & $7.36\times10^{-33}$ & $4.23\times10^{-31}$ \\ \hline 

\end{tabular}
\vspace{0.6in}
\caption{Comparison between average neutrino(antineutrino)-nucleon charged current cross section in cm$^{2}$ according to the numerical results obtained in Ref. \protect\cite{FIOREJKPPPR12546} (2nd column), the asymptotic approximation method of Ref. \protect\cite{FIOREJKPPPR12} (3rd column), our analytical evaluation (4th column) obtained using eq. (\ref{eqn:ch6eq44analytical}) as well as eq. (\ref{eqn:ch6eq46analytical}) and our numerical evaluation (5th column) obtained using eq. (\ref{eqn:ch6eq44numerical}) as well as eq. (\ref{eqn:ch6eq46numerical}).}
\label{table:ch6tab6}
\end{center}
\end{table}

\begin{table}
\begin{center}
\vspace{-0.3in}
\begin{tabular}{|c|c|c|}\hline
\multicolumn{3}{|c|}{\rule[-.3cm]{0mm}{1.4cm}\bfseries  \large{\bf{Upper bounds on cross-section }}} \\ \hline {\rule[-.3cm]{0mm}{1cm}} {$E_{\nu}$ (GeV)}  & Rice \cite{Krai,IKRAVCHENCO}  &  Agasa \cite{YOSHIDAS} \\ \cline{2-3} &  \hspace{-.8cm} FKRT \cite{ZFODORSDKARHTU}  \vline  PJ \cite{RJPROPAJOHN} &  \hspace{-.8cm} FKRT \cite{ZFODORSDKARHTU} \vline   PJ \cite{RJPROPAJOHN}  \\ \hline {\rule[-.3cm]{0mm}{1cm}}
$10^{10}$ 	& $1.2\times10^{-30}$	  \vline $2.8\times10^{-31}$  & $1.3\times10^{-28}$  \vline$2.4\times10^{-29}$ \\ \hline {\rule[-.3cm]{0mm}{1cm}}
$10^{10.5}$	& $3.6\times10^{-30}$	  \vline $7.2\times10^{-31}$  & No  result.\,\,\,  \vline   $1.4\times10^{-28}$ \\ \hline {\rule[-.3cm]{0mm}{.8cm}}
$10^{11}$	& $1.9\times10^{-29}$	  \vline $3.8\times10^{-30}$  & No result. \vline       No result.\\ \hline 

\end{tabular}
\vspace{0.6in}
\caption{Results of the upper bounds on cross-section cm$^{2}$ in Rice \protect\cite{Krai,IKRAVCHENCO} and Agasa \protect\cite{YOSHIDAS} experiment as per the estimation of neutrino flux of Ref. \protect\cite{ZFODORSDKARHTU} and Ref. \protect\cite{RJPROPAJOHN}.}
\label{table:ch6tab12345}
\end{center}
\end{table}

\begin{table}
\begin{center}
\begin{tabular}{|c|c|c|c|c|}\hline
\multicolumn{5}{|c|}{\rule[-.3cm]{0mm}{1.4cm}\bfseries  \large{\bf{Our results versus Upper bounds on cross-section }}} \\ \hline {\rule[-.3cm]{0mm}{1.2cm}}
$E_{\nu}$\,\,(GeV)  & Analytical\,  & Numerical \, &  Ref \cite{RCMI1}\, &  Ref \cite{BELBLOCKMCAN}\, \\ \hline {\rule[-.3cm]{0mm}{1cm}}
$10^{10}$ 	& $1.88\times10^{-33}$	     & $2.47\times10^{-32}$  &$4.35\times10^{-32}$   & $2.24\times10^{-32}$  \\ \hline {\rule[-.3cm]{0mm}{1cm}}
$10^{10.5}$	& $2.32\times10^{-33}$	     & $3.28\times10^{-32}$  & No  result.           & No  result.  \\ \hline {\rule[-.3cm]{0mm}{.8cm}}
$10^{11}$	& $2.88\times10^{-33}$	     & $5.97\times10^{-32}$  & $1.02\times10^{-31}$  & $3.66\times10^{-32}$  \\ \hline 

\end{tabular}
\vspace{0.6in}
\caption{Our analytical and numerical results alongwith the results of the Ref. \protect\cite{RCMI1} and Ref. \protect\cite{BELBLOCKMCAN} to be compared with the experimental upper bounds on cross-section in cm$^{2}$ as given in Table \ref{table:ch6tab12345}.}
\label{table:ch6tab10100}
\end{center}
\end{table}

\begin{table}
\begin{center}
\begin{tabular}{|c|c|c|}\hline
\multicolumn{3}{|c|}{\rule[-.3cm]{0mm}{1.4cm}\bfseries  \large{\bf{Sensitivity of PAO}}} \\ \hline {\rule[-.3cm]{0mm}{1.2cm}}
$E_{\nu}$\,\,(GeV)  & FKRT \cite{ZFODORSDKARHTU}\,  & PJ \cite{RJPROPAJOHN}  \\ \hline {\rule[-.3cm]{0mm}{1cm}}
$10^{10}$ 	& $1.0\times10^{-30}-1.9\times10^{-30}$	     & $3.1\times10^{-31}-5.6\times10^{-31}$    \\ \hline {\rule[-.3cm]{0mm}{1cm}}
$10^{10.5}$	& $4.2\times10^{-30}-8.7\times10^{-30}$	     & $1.1\times10^{-30}-2.1\times10^{-30}$     \\ \hline {\rule[-.3cm]{0mm}{.8cm}}
$10^{11}$	& $2.7\times10^{-29}-7.8\times10^{-29}$	     & $6.2\times10^{-30}-1.5\times10^{-29}$    \\ \hline 

\end{tabular}
\vspace{0.6in}
\caption{Sensitivity of Pierre Auger Observatory (PAO) at 95\% confidence limit in $cm^2$. This is derived with the assumption that no deeply developing showers above Standard Model background has been observed in 5 years of operation. This can be compared with the results given in Table \ref{table:ch6tab10100}.}
\label{table:ch6tabGermany}
\end{center}
\end{table}

\begin{table}
\begin{center}
\vspace{-0.3in}
\begin{tabular}{|c|c|c|c|c|}\hline
\multicolumn{5}{|c|}{\rule[-.3cm]{0mm}{1.4cm}\bfseries  \large{\bf{Energy cum total neutrino-nucleon interaction cross-section }}} \\ \hline {\rule[-.3cm]{0mm}{1.2cm}}
$E_{\nu}$\,\,(GeV$^2$)  &  $\sigma_{total}^{\nu N}$, Ref. \cite{RAHULBASU} & $\sigma_{total}^{\nu N}$, Ref. \cite{RCMI111} &  $\sigma_{total}^{\nu N}$, analytical &  $\sigma_{total}^{\nu N}$, numerical   \\ \hline {\rule[-.3cm]{0mm}{1cm}}

$10^{7}$	& $2.48\times10^{-33}$	     &$2.72\times10^{-33}$ & $5.55\times10^{-34}$ & $1.35\times10^{-33}$ \\ \hline {\rule[-.3cm]{0mm}{.8cm}}
$10^{8}$	& $5.59\times10^{-33}$	     &$6.29\times10^{-33}$ & $8.28\times10^{-34}$ & $3.65\times10^{-33}$ \\ \hline {\rule[-.3cm]{0mm}{.8cm}}
$10^{9}$	& $1.26\times10^{-32}$	     &$1.45\times10^{-32}$  & $1.24\times10^{-33}$ & $9.70\times10^{-33}$ \\ \hline {\rule[-.3cm]{0mm}{.8cm}}
$10^{10}$	& $2.83\times10^{-32}$	     &$3.34\times10^{-32}$  & $1.88\times10^{-33}$ & $2.47\times10^{-32}$ \\ \hline {\rule[-.3cm]{0mm}{.8cm}}
$10^{11}$	& $6.37\times10^{-32}$	     &$7.71\times10^{-32}$ & $2.88\times10^{-33}$ & $5.97\times10^{-32}$ \\ \hline
\end{tabular}
\vspace{0.6in}
\caption{Comparison between the neutrino-nucleon total cross section in cm$^{2}$ according to the numerical results obtained in Ref. \protect\cite{RAHULBASU} (2nd column), the numerical method of Ref. \protect\cite{RCMI111} (3rd column), our analytical evaluation (4th column) obtained using eq. (\ref{eqn:ch6eq44analytical}) as well as eq. (\ref{eqn:ch6eq45analytical}) and our numerical evaluation (5th column) obtained using eq. (\ref{eqn:ch6eq44numerical}) as well as 
eq. (\ref{eqn:ch6eq45numerical}).}
\label{table:ch6tab7}
\end{center}
\end{table}

\begin{table}
\begin{center}
\vspace{-0.3in}
\begin{tabular}{|c|c|c|c|}\hline
\multicolumn{4}{|c|}{\rule[-.3cm]{0mm}{1.75cm}\bfseries  \large{\bf{Energy versus $\displaystyle{\left(\frac{\sigma_{CC}^{\nu N}}{\sigma_{CC}^{\overline{\nu}N}}\right)}$}}} \\ \hline {\rule[-.3cm]{0mm}{1.50cm}}
$E_{\nu(\overline{\nu})}$\,\,(GeV$^2$)  &   analytical &  numerical &  Ref \cite{ACSSS} \\ \hline {\rule[-.3cm]{0mm}{1.2cm}}

$10$            & $4.36$     & $2.30$  & NO.\\ \hline {\rule[-.3cm]{0mm}{1cm}}
$10^{2}$	& $2.19$     & $2.06$  & $2.06$\\ \hline {\rule[-.3cm]{0mm}{.8cm}}
$10^{3}$	& $1.75$     & $1.94$  & $1.82$\\ \hline {\rule[-.3cm]{0mm}{.8cm}}
$10^{4}$	& $1.47$     & $1.65$  & $1.52$\\ \hline {\rule[-.3cm]{0mm}{.8cm}}
$10^{5}$	& $1.26$     & $1.27$  & $1.19$\\ \hline {\rule[-.3cm]{0mm}{.8cm}}
$10^{6}$	& $1.10$     & $1.07$  & $1.05$\\ \hline {\rule[-.3cm]{0mm}{.8cm}}
$10^{7}$	& $1.03$     & $1.02$  & No.   \\ \hline {\rule[-.3cm]{0mm}{.8cm}}
$10^{8}$	& $1.01$     & $1.00$  & No.   \\ \hline {\rule[-.3cm]{0mm}{.8cm}}
$10^{9}$	& $1.00$     & $1.00$  & No.   \\ \hline {\rule[-.3cm]{0mm}{.8cm}}
$10^{10}$	& $1.00$     & $1.00$  & No.   \\ \hline {\rule[-.3cm]{0mm}{.8cm}}
$10^{11}$	& $1.00$     & $1.00$  & No.   \\ \hline {\rule[-.3cm]{0mm}{.8cm}}
$10^{12}$	& $1.00$     & $1.00$  & No.   \\ \hline

\end{tabular}
\vspace{0.6in}
\caption{The ratio of neutrino-nucleon charged current cross-section and antineutrino-nucleon charged current cross-section evaluated by our analytical and numerical method is plotted as a function of neutrino (antineutrino) energy $\left(E_{\nu(\overline{\nu})}\right)$. This is compared with the ratio obtained from Ref \protect\cite{ACSSS} }
\label{table:ch6tab8}
\end{center}
\end{table}
\newpage
In Figure  \ref{fig:ch6fig1}, we plot our analytical LO results (Magenta coloured dash dot line) for $\sigma_{CC}^{\nu N}\,(cm^{2})$  eq. (\ref{eqn:ch6eq44analytical}) (Table \ref{table:ch6tab1}). We then plot exact numerical LO results (Blue coloured dash line) calculated by us eq. (\ref{eqn:ch6eq44numerical}) (Table \ref{table:ch6tab3}). In the same figure, we also plot NLO results of Ref. \cite{GKR} eq. (\ref{eqn:ch6eq44}) (Pink coloured dot line) and that of Ref. \cite{RCMI111} (Green coloured solid line) eq. (\ref{eqn:ch6eq48}) as well as Ref. \cite{RCMI1} eq. (\ref{eqn:ch6eq52}) (Wine coloured dash dot dot line). 

In Figure  \ref{fig:ch6fig2}, we compare our LO predictions (Magenta coloured dash dot line) for $\sigma_{NC}^{\nu N}\,(cm^{2})$ eq. (\ref{eqn:ch6eq45analytical}) (Table \ref{table:ch6tab1}), with the exact numerical LO results calculated by us  eq. (\ref{eqn:ch6eq45numerical}) (Blue coloured dash line) (Table \ref{table:ch6tab3}). The NLO results of Ref. \cite{GKR} eq. (\ref{eqn:ch6eq45}) (Pink coloured dot line) and that of Ref. \cite{RCMI111} (Green coloured solid line) eq. (\ref{eqn:ch6eq49}) as well as Ref. \cite{RCMI1}  eq. (\ref{eqn:ch6eq53}) (Wine coloured dash dot dot line) are then plotted to have a relative comparison.

Similarly in Figure  \ref{fig:ch6fig3}, we compare our LO results for $\sigma_{CC}^{\overline{\nu} N}\,(cm^{2})$ (Magenta coloured dash dot line) eq. (\ref{eqn:ch6eq46analytical}) (Table \ref{table:ch6tab2}) with the exact numerical LO results calculated by us eq. (\ref{eqn:ch6eq46numerical}) (Blue coloured dash line) (Table \ref{table:ch6tab4}), the NLO results of Ref. \cite{GKR} eq. (\ref{eqn:ch6eq46}) (Pink coloured dot line), Ref. \cite{RCMI111} (Green coloured solid line) eq. (\ref{eqn:ch6eq50}) as well as Ref. \cite{RCMI1} eq. (\ref{eqn:ch6eq55}) (Wine coloured dash dot dot line).

In  Figure \ref{fig:ch6fig4}, we plot our LO results for $\sigma_{NC}^{\overline{\nu} N}\,(cm^{2})$ (Magenta coloured dash dot line) eq. (\ref{eqn:ch6eq47analytical})(Table \ref{table:ch6tab2}) and then compare with the exact numerical LO results calculated by us eq. (\ref{eqn:ch6eq47numerical}) (Blue coloured dash line) (Table \ref{table:ch6tab4}). For relative comparison, the NLO results of Ref. \cite{GKR} eq. (\ref{eqn:ch6eq47}) (Pink coloured dot line), Ref. \cite{RCMI111} (Green coloured solid line) eq. (\ref{eqn:ch6eq51}) as well as Ref. \cite{RCMI1} eq. (\ref{eqn:ch6eq56}) (Wine coloured dash dot dot line) are also plotted. 

   In Figure \ref{fig:ch6fig5}, we compare the value of $\overline{\sigma}_{CC}^{\nu N}$ obtained by using our analytical (Magenta coloured dash dot line) 
(using eq. (\ref{eqn:ch6eq44analytical}) as well as eq. (\ref{eqn:ch6eq46analytical})) and numerical techniques (Wine coloured dot line) (using eq. (\ref{eqn:ch6eq44numerical}) as well as eq. (\ref{eqn:ch6eq46numerical})) with the numerical results obtained in Ref. \cite{FIOREJKPPPR12546} (Blue coloured solid line), the asymptotic approximation method of Ref. \cite{FIOREJKPPPR12} (Red coloured dash line) (Table \ref{table:ch6tab6}). It is pertinent to note that the factor $\overline{\sigma}_{CC}^{\nu N}$ is defined as: $\overline{\sigma}_{CC}^{\nu N}=\displaystyle{\frac{\sigma_{CC}^{\nu N} + \sigma_ {CC}^{\overline{\nu} N}}{2}}$.

In Figure \ref{fig:ch6fig6}, we compare between the neutrino-nucleon total cross section in cm$^{2}$ according to the numerical results obtained in Ref. \cite{RAHULBASU} (Blue coloured solid line), the numerical method of Ref. \cite{RCMI111} (Magenta coloured dash line), our analytical evaluation (Olive coloured dot line) and our numerical evaluation based on GRV94 distribution (Red coloured dash dot dot line) (Table \ref{table:ch6tab7}).  

In Figure \ref{fig:ch6fig7}, the ratios of neutrino-nucleon charged current cross-section and antineutrino-nucleon charged current cross-section $\displaystyle{\left(\frac{\sigma_{CC}^{\nu N}}{\sigma_{CC}^{\overline{\nu}N}}\right)}$ for both of our analytical (Blue coloured solid line) and numerical methods (Red coloured dash line) respectively are plotted as a function of neutrino (antineutrino) energy $\left(E_{\nu(\overline{\nu})}\right)$. The ratio obtained from Ref \cite{ACSSS} (Magenta coloured dash dot line) is compared with our analytical and numerical results (Table \ref{table:ch6tab8}).

\subsection{Comparative study and limitations:}
\label{ch6:Comparative study and limitations}
Let us make a comparative study of the Figures (\ref{fig:ch6fig1}-\ref{fig:ch6fig6}). As expected and clearly demonstrated in these figures, our analytical result  indicates a sharp rise for $E_{\nu}\le 10^{3}$ GeV. This is due to the absence of propagator effect in low energy regime \cite{RCMI}. For $E_{\nu}\ge 10^{3}$ GeV, the dampening due to the propagator effect takes place. Here we find our numerically determined result in sufficient agreement with the results obtained by the above-mentioned authors \cite{RAHULBASU,RCMI1,GKR,RCMI111,FIOREJKPPPR12546}. As far as our analytical result is concerned, we find it to match better with other results at lower end of the energy spectrum ($10^{5}\le E_{\nu(\overline{\nu})}\le 10^{8}$) of the UHE neutrino rather than the higher end ($10^{8}\le E_{\nu(\overline{\nu})}\le 10^{12}$). At the higher end, we observe  dampening of the cross-section than what is expected from the  other numerical results etc. mentioned beforehand. The quantitative difference at the higher end is presumably due to the following factors:

(1) We have not extrapolated MRST distribution for values of $x$ below $10^{-5}$, assuming that our approximate analytical solutions of DGLAP equations for singlet structure function $F_{2}^{S}(x,t)$ as well as non-singlet structure function $xF_{3}$ remains valid in the UHE regime.

(2) We have considered solutions at tree level. Replacing LO distributions by NLO distributions and with the appropriate addition of convolutions with the fermionic Wilson coefficient $C_q$ and the NLO contribution $C_g \otimes g$ \cite{FURPETRONZIO1}, we may expect somewhat better results.

(3) We have neglected the finite $x$ corrections coming from the higher derivatives of $\displaystyle{\frac{\partial F^{NS}(x,t)}{\partial x}}$ in Taylor approximation of DGLAP equations in case of $t$-evolution of $xF_{3}$ at one loop level \cite{DKCPKD}. Similarly, we have neglected the finite $x$ corrections coming from the higher derivatives of $\displaystyle{\frac{\partial F_{2}^{S}(x,t)}{\partial x}}$ and $\displaystyle{\frac{\partial G(x,t)}{\partial x}}$ in Taylor approximation of DGLAP equations in case of $t$-evolution of $F_{2}$ at one loop level \cite{Campus}.

(4) We have considered the contribution due to light quarks only and neglected the contribution due to heavy quarks. For $E_{\nu}\ge 10^{8}$ GeV, preferably for $E_{\nu}\rightarrow 10^{12}$ GeV,  NLO massive $t\bar{b}$ contribution will introduce an ${\cal{O}}(\alpha_s)$ correction of about 20\% to  the LO contribution of $\sigma_{\rm tot}^{\nu N}$.
               
We find that the blue coloured dashed line drawn on the basis of our numerical method has got the almost same slope as that of the other lines drawn on the basis other numerical methods as seen in Figures (\ref{fig:ch6fig1}-\ref{fig:ch6fig4}), proving the excellence of our numerical results. A little lateral shift is presumably because of the following reasons:

(1) We have done calculations at leading order involving LO parton distributions, whereas the calculations done by other authors at next-to-leading order are based on NLO parton distributions. Obviously a better result is expected at NNLO level. 

(2) It has been found that there had been a $10\%$ discrepancies between GRV 94 distribution \cite{GRVARXIV55A} adopted by us for numerical calculation and the precision measurements at HERA done at a later stage (in 1996 and 1997). This implies that GRV 94 distribution has restricted accuracy margin. So Gluck et al realised the necessity of the fine tuning of the input parameter $\mu$ and/or $f(x, \mu^{2})$ both at LO and NLO level which they did and removed these $10\%$ discrepencies by providing the GRV 98 parton distributions \cite{GRVARXIV55}. It is pertinent to remind that  NLO calculations of Gluck et al were based mainly on GRV 98 parton distributions, which was almost free from any uncertainty of radiative predictions because of timely fine tuning. For example, they fine-tuned $\mu_{LO}^{2}$ from the value $0.23$ GeV$^{2}$ to $0.26$ GeV$^{2}$ at one loop level. Similarly there was a fine tuning at two loop level from from the value $0.34$ GeV$^{2}$ to $0.40$ GeV$^{2}$ \cite{GRVARXIV55A,GRVARXIV55}.

(3) GRV 94 predictions adopted by us are valid in the small $x$ region $10^{-5}\le x < 1$. We did our numerical calculations assuming the validity of these distributions in the UHE regime i.e. $10^{-12}\le x < 10^{-5}$, which is obviously not true. Uncertainty and error are definite to crop up (however small it may be) due to the this non-extrapolation. It is reminded that Gluck et al later extended their perturbatively stable parameter–free dynamical predictions to the extremely small-$x$ region $10^{-9}\le x < 10^{-5}$ \cite{GRVARXIV55}.

(4) We have not taken into consideration the contribution due to heavy quarks.

\subsection{Upper bounds on the neutrino-nucleon inelastic cross-sections:}
Model independent upper bounds on the neutrino-nucleon inelastic cross-sections derived from the RICE \cite{Krai,IKRAVCHENCO} and AGASA \cite{YOSHIDAS} collaboration search results by exploiting the cosmogenic neutrino flux estimates of Ref. \cite{ZFODORSDKARHTU} and Ref. \cite{RJPROPAJOHN} are already known. This is shown in Table \ref{table:ch6tab12345} \cite{ANCHORFDSKARIWTUH}. Such an upper bound arises due to non-observation of events triggered off by UHE neutrinos as reported by RICE and AGASA. We compare our analytical as well as numerically determined results for the Ultra High Energy neutrino-nucleon cross-section in leading order with the above-mentioned upper bounds at three different neutrino energies $E_\nu = 10^{10}, 10^{10.5}, 10^{11}$ GeV.  We find that our results do not violate the experimental upper bounds imposed on the neutrino-nucleon interaction cross-section and are well within it as clearly seen in  Table \ref{table:ch6tab12345}  and Table \ref{table:ch6tab10100}. In Table \ref{table:ch6tabGermany}, Sensitivity of Pierre Auger Observatory (PAO) at 95\% confidence limit in $cm^2$ is shown \cite{ANCHORFDSKARIWTUH}. This has been derived with the assumption that no deeply developing showers above Standard Model background has been observed in 5 years of operation. This can be compared with the results given in Table \ref{table:ch6tab10100}. 

In Figure \ref{fig:Trombay}, we plot our analytical result (Wine coloured dot line) and numerical (exact) result (Magenta coloured dash dot line) for neutrino-nucleon cross-section derived from Standard Model. This is compared with the upper bounds on the neutrino-nucleon cross-sections derived from the RICE  collaboration search results by exploiting the cosmogenic neutrino flux estimates of FKRT \cite{ZFODORSDKARHTU} (Red coloured Solid line) and PJ \cite{RJPROPAJOHN} (blue coloured dash line).

\subsubsection{Beyond the standard model scenario:}
As per Beyond the Standard model scenario, neutrinos with energies around $10^{10.5}$ GeV may behave as strongly interacting as protons, which has the capacity of inducing vertical air showers at high altitudes in the atmosphere. Since there is a good match of cosmogenic neutrino flux with the UHE cosmic ray spectrum above Greison-Zatsepin-Kuzmin (GZK) energy ($E_{GZK} \approx 10^{10.9}$ GeV), so one can appreciate this 'strongly interacting' characteristics of cosmogenic neutrinos which can solve the GZK puzzle. If the strongly interacting neutrino scenario is considered for $E_{\nu}\succeq 10^{11}$ GeV, which leads to the generation of vertical showers, then non-observation of quasi-horizontal air showers by AGASA would not be able to constrain the neutrino-nucleon cross-section the way it did.

\section{Comments and conclusions:}
\label{secchap6:Comments and conclusions}
To conclude, we have shown that our analytical evaluation of neutrino(antineutrino)-nucleon cross-section tallies well in selected part of  of the neutrino(antineutrino) energy spectrum. Particularly it tallies well with the result obtained  at the lower end of the energy spectrum, but at the higher end there is  dampening much more than what is expected due to propagator effect in case of numerical solutions by other authors \cite{RAHULBASU,RCMI1,GKR,FIOREJKPPPR12,RCMI111,FIOREJKPPPR12546}. The reason for such differences at higher end of the neutrino(antineutrino) energy spectrum has been explained in subsection \ref{ch6:Comparative study and limitations}. On the other hand, our LO numerical result tallies quite well with the NLO results of other authors both at the lower as well as higher ends of the neutrino(antineutrino) energy spectrum. The possible reasons for minute lateral shift has been explained adequetly in subsection \ref{ch6:Comparative study and limitations}.

\section{Acknowledgement:}
One of the authors (P.K.Dhar) offers his gratitude to  Dr. Mike Whalley of Durham University, U.K.,  Dr. Pedro Jimenez-Delgado of the Institute for Theoretical Physics, University of Zurich, Switzerland and Prof. A. D. Martin of Durham University, U.K. for some very useful interactions about parton distributions, to Prof. Raj Gandhi of Harish Chandra Research Institute, Allahabad, India, Late Prof. Rahul Basu of Institute of Mathematical Sciences, Chennai, India and Prof. Pijushpani Bhattacharjee of Saha Institute of Nuclear Physics, Kolkata, India, Prof. Naba K Mondal, Department of High Energy Physics, Tata Institute of Fundamental Research, Mumbai, India  for useful discussions on UHE neutrino-nucleon interaction and to Prof. M. Guzzi, Department of Physics, University of Crete, Greece for some useful informations.

\end{document}